\documentclass[12pt]{article}

\usepackage{epsfig,amsmath,cite}

\newcommand{\mupi}{\mu_\pi^2}
\newcommand{\mug}{\mu_G^2}
\newcommand{\rd}{\rho_D^3}
\newcommand{\rls}{\rho_{LS}^3}
\newcommand{\as}{\alpha_s}

\newcommand{\GeV}{\,\mbox{GeV}}

\newcommand\gsim{\mathop{\mbox{\vbox{\hbox{$>$} \vskip -9pt \hbox{$\sim$}
             \vskip -3pt  }}}}
\newcommand\lsim{\mathop{\mbox{\vbox{\hbox{$<$} \vskip -9pt \hbox{$\sim$}
             \vskip -3pt  }}}}

\def \be{\begin{equation}}
\def \ee{\end{equation}}
\newcommand{\bea}{\begin{eqnarray}}
\newcommand{\eea}{\end{eqnarray}}

\begin{document}
\begin{titlepage}

\vskip 3cm

\centerline{\LARGE\bf\boldmath Inclusive semileptonic fits,
}
\vskip 2mm
\centerline{\LARGE\bf\boldmath  heavy quark masses, and $V_{cb}$ 
}
\vskip 2cm

\begin{center}
{\bf 
  Paolo Gambino \\[3mm]
\it   Dip.\ di  Fisica, Univ.\ di Torino, \& INFN  Torino,
I-10125 Torino, Italy\\[2mm]
\rm and \\[4mm]
\rm \bf Christoph Schwanda\\[3mm]
\it Institut f\"ur Hochenergiephysik, Wien, Austria
}
\end{center}

\vskip 2cm

\begin{abstract}
We perform  global fits to the moments of semileptonic $B$ decay distributions  and extract $|V_{cb}|$,  the heavy quark masses,  and the non-perturbative parameters of 
the heavy quark expansion. We include NNLO perturbative corrections and recent 
determinations of the charm mass, and discuss how they improve the 
precision of the global fit. In particular, 
using the $m_c$ determination of Ref.\cite{mcmb_karlsruhe} we get $m_b^{kin}=4.541(23)\GeV$ and $|V_{cb}|=(42.42\pm 0.86)\times 10^{-3}$.
We also discuss the implications of the new fits for the normalization of rare $B$ decays, the 
zero-recoil sum rule in $B\to D^* \ell\nu$,
and the inclusive determination of   $|V_{ub}|$.
\end{abstract}

\end{titlepage}


\section{Introduction}
The CKM matrix elements $V_{cb}$ and  $V_{ub}$ are important  ingredients in the 
analyses of CP violation in the Standard Model and in the search for new physics in 
flavor violating processes. For instance, 
 the absolute value of their ratio gives one of the sides of the unitarity triangle, and  the
$\varepsilon_K$ constraint on $\bar\rho$ and $\bar\eta$    is very sensitive to the precise value of 
$|V_{cb}|$, see \cite{UTfit} for recent analyses.  The determination of $V_{cb}$ and  $V_{ub}$ from inclusive semileptonic $B$ decays is based on an Operator Product Expansion (OPE)
that allows us to express  the  widths and the first moments of the kinematic distributions of $B\to X_{u,c} \ell\nu$ as double expansions in
$\as$ and $\Lambda_{\rm QCD}/m_b$.   The leading terms in these double expansions are given by the free $b$ quark decay, and 
the first corrections are $O(\alpha_s)$ and $O(\Lambda^2_{\rm QCD}/m_b^2)$\cite{1mb2}.

The relevant parameters of the double expansions are the heavy quark masses $m_b$ 
and  $m_c$, the strong 
coupling $\alpha_s$, and the $B$-meson matrix elements of  local operators of growing 
dimension: $\mupi$ and $\mug$ at  
$O(1/m_b^2)$, $\rd$  and  $\rls$ at $O(1/m_b^3)$, etc.  The latter can be constrained by 
various moments of the lepton energy and hadron mass distributions of $B\to X_c \ell \nu$ 
that have been   measured with good accuracy at the $B$-factories, as well as at CLEO, DELPHI, CDF.
The total semileptonic width can then be employed to extract  $|V_{cb}|$. The situation is 
less favorable in the case of $|V_{ub}|$, where the total rate is much more difficult to access experimentally,
but the results of the semileptonic fits are  crucial in that case as well,  see \cite{Antonelli:2009ws} for 
a review.  This strategy has been rather successful and has allowed for a $\sim 2\%$ determination of
$V_{cb}$ and for a  $\sim 5\%$ determination of
$V_{ub}$ from inclusive decays \cite{HFAG}. 

Complementary studies of exclusive  decays and 
non-perturbative calculations of the relevant form-factors have also progressed considerably,  reaching a similar level of accuracy.  Unfortunately,  
a  $\sim2\sigma$ discrepancy persists between the most precise determinations of $|V_{cb}|$:
the inclusive one and the one based on $B\to D^* \ell\nu$ at zero recoil and a lattice calculation of the form-factor \cite{HFAG,lattice}.  However,
the zero-recoil form-factor estimate based on heavy quark sum rules leads to $|V_{cb}|$ in good agreement with the inclusive result \cite{long}.  A stronger discrepancy between the inclusive and exclusive determinations occurs in the case of $|V_{ub}|$ \cite{HFAG}.

The importance of a precise and reliable extraction of $|V_{cb}|$ and of the inputs for
the inclusive $|V_{ub}|$ analysis has motivated us to critically re-examine the procedure used in the semileptonic fits almost a decade after the first comprehensive studies \cite{babarfit,1S,BF}. There are three relevant issues in this respect:  $i)$ the theoretical uncertainties and how they are implemented in the fit, $ii)$ the
inclusion of additional constraints on the parameters, $iii)$ the need to update the theoretical predictions to NNLO.

For what concerns the
theoretical uncertainties,  we have already observed \cite{Gambino:2011fz}
a marked dependence of the results on the ansatz for the correlations among theoretical 
uncertainties at different values of the cut on the lepton energy. In this paper we discuss 
different options and compare the results of the fits performed accordingly.

Regarding the inclusion of 
additional constraints from other processes, we recall that  the semileptonic moments are sensitive to a linear combination of $m_c$ and $m_b$, but cannot resolve the individual 
masses with good accuracy. To improve the accuracy of the fit, the moments of the photon energy  in  $B\to X_s \gamma$ have generally been employed,
but in the last few years
quite precise determinations of the heavy quark masses by completely different methods ($e^+e^-$ sum rules, lattice QCD etc.) have 
become  available, see {\it e.g.} \cite{mcmb_karlsruhe,masseslat,mc_hoang,mb_hoang,simula}.  Radiative moments remain very interesting, but  they are not competitive with the charm mass determinations and are in 
principle subject to additional $O(\Lambda_{QCD}/m_b)$ effects, which have not yet been estimated \cite{paz}.  
Here we will discuss  the inclusion of the most recent 
heavy quark mass determinations in the semileptonic fit. As we will see, these precise external inputs reduce the dependence on the assumptions regarding the theoretical correlations.

Recent progress has made it necessary to  update  the theoretical predictions used in the fits.  The 
NNLO $O(\as^2)$ calculation has been completed \cite{czarnecki-pak,melnikov,melnikov2} 
and implemented  \cite{NNLO} in a code that has been used in
preliminary analyses by the Heavy Flavor Averaging Group \cite{HFAG}. This code extends and improves on  the kinetic scheme calculation of Ref.\cite{btoc}, used in \cite{babarfit,BF},
and forms the basis of the present work. The NNLO accuracy is a necessary 
prerequisite for using the precise heavy quark mass constraints we have just mentioned. 
Concerning  the perturbative corrections to the 
Wilson coefficients of power-suppressed operators, the $O(\alpha_s\mu_\pi^2/m_b^2)$ have been computed in \cite{Becher:2007tk,Alberti:2012dn}. However, the $O(\alpha_s\mu_G^2/m_b^2)$ corrections are not yet available and here
we will   not  include any  $O(\alpha_s/m_b^2)$ corrections. 
Concerning  higher order power corrections, the 
 $O(1/m_b^4)$
 and $O(1/m_Q^5)$ effects  have been computed \cite{Mannel:2010wj}.
The main problem here is a proliferation of non-perturbative  parameters 
that  cannot all be fitted  from experiment. In Ref.~\cite{Mannel:2010wj}
they were estimated in the ground state saturation approximation, leading to 
a final $O(1/m_Q^{4,5})$ 
effect on $|V_{cb}|$ of  $+0.4\%$.  
This may well be the order of magnitude of higher order power corrections, but
further investigations are necessary,  also because sizable effects beyond  the ground state 
saturation approximation have been found  \cite{long}. Here, we include  
only effects up to $O(1/m_b^3)$ \cite{1mb3}.

The paper is organized in the following way. We first discuss the relevant observables and 
list all the moment measurements included in our fits. Then, in Section 3, we explain our estimate of theoretical errors and discuss several options for their correlations. 
In Section 4 we consider the inclusion of independent constraints on $m_c$ and $m_b$ in 
the fits and study their impact on the results; then,  in Section 5, we consider a few relevant applications of the fits. Finally, Section 6 contains a summary of our results.

\section{Observables included in the fits}
The first few moments of the charged lepton energy spectrum in inclusive  $b\to c\ell \nu$ decays are experimentally measured with high precision ---
better than 0.2\% in the case of the first moment. 
At the $B$-factories a lower cut on the lepton energy, $E_\ell \ge E_{cut}$, is  applied to suppress the  background. Experiments measure  the moments at different values of $E_{cut}$, which provides additional information as the cut dependence is also a function of the OPE parameters. The relevant quantities are therefore
\begin{equation}
\langle E_\ell^{n}\rangle_{E_\ell > E_{cut}} = \frac{\int_{E_{cut}}^{E_{max}} d E_\ell \ E_\ell^n \ \frac{d\Gamma}{d E_\ell}}
{\int_{E_{cut}}^{E_{max}} d E_\ell \ \frac{d\Gamma}{d E_\ell}}\ ,
\end{equation}
which are measured for $ n $ up to 4, as well as the ratio $R^*$ between the rate with and without a cut
\begin{equation}
R^* (E_{cut})= \label{eq:Rstar}
\frac{\int_{ E_{cut}}^{E_{max}} d E_\ell \ \frac{d\Gamma}{d E_\ell}}
{\int_0^{E_{max}} d E_\ell \ \frac{d\Gamma}{d E_\ell}}\ .
\end{equation}
This quantity is needed to relate the actual measurement of the rate with a cut to the total rate, from which one conventionally extracts $|V_{cb}|$. 
Since the physical information that can be extracted from the first three linear moments 
is highly correlated, it is more convenient to study the central moments, namely the variance 
and asymmetry of the lepton energy distribution. In the following we will consider only $R^*$ 
and 
\be
\ell_1(E_{cut})=\langle E_\ell \rangle_{E_\ell > E_{cut}}, \quad\quad \quad
\ell_{2,3}(E_{cut})=\langle \left(E_\ell - \langle E_\ell \rangle \right)^{2,3} \rangle_{E_\ell > E_{cut}}\,.
\ee
Similarly, in the case of the moments of the hadronic invariant mass distribution, 
we consider 
\be
h_1(E_{cut})=\langle M_X^2\rangle_{E_\ell > E_{cut}}, \quad \quad
h_{2,3}(E_{cut})=\langle (M_X^2-\langle M_X^2\rangle)^{2,3}\rangle_{E_\ell > E_{cut}} .
\ee
\begin{table}[t]
  \begin{center} \begin{tabular}{|c|c|c|c|}
    \hline 
     & experiment & values of $E_{cut}(\rm GeV)$ &Ref.\\ \hline
$R^*$ & BaBar&  0.6,\ 1.2,\ 1.5   &\cite{Aubert:2009qda,Aubert:2004td}\\
$\ell_{1}$ & BaBar&  0.6,\ 0.8,\ 1,\ 1.2,\ 1.5  &\cite{Aubert:2009qda,Aubert:2004td}\\
$\ell_{2}$ & BaBar&  0.6,\ 1,\ 1.5  &\cite{Aubert:2009qda,Aubert:2004td}\\
$\ell_{3}$ & BaBar&  0.8,\ 1.2  &\cite{Aubert:2009qda,Aubert:2004td}\\
 $h_{1}$ & BaBar& 0.9,\ 1.1,\ 1.3,\ 1.5 & \cite{Aubert:2009qda}\\
 $h_{2}$ & BaBar& 0.8,\ 1,\ 1.2,\ 1.4 & \cite{Aubert:2009qda}\\
 $h_{3}$ & BaBar& 0.9,\ 1.3 & \cite{Aubert:2009qda}\\
 $R^*$ & Belle&  0.6,\ 1.4   &\cite{Urquijo:2006wd}\\
$\ell_{1}$ & Belle&  1,\ 1.4  &\cite{Urquijo:2006wd}\\
$\ell_{2}$ & Belle&  0.6,\ 1.4  &\cite{Urquijo:2006wd}\\
$\ell_{3}$ & Belle&  0.8,\ 1.2  &\cite{Urquijo:2006wd}\\
 $h_{1}$ & Belle& 0.7,\ 1.1,\ 1.3,\ 1.5 & \cite{Schwanda:2006nf}\\
 $h_{2}$ & Belle& 0.7,\ 0.9,\ 1.3 & \cite{Schwanda:2006nf}\\
 $h_{1,2}$ & CDF& 0.7 & \cite{Acosta:2005qh}\\
  $h_{1,2}$ & CLEO& 1,\ 1.5 & \cite{Csorna:2004kp}\\
 $\ell_{1,2,3}$ & DELPHI&  0  &\cite{delphi}\\
 $h_{1,2,3}$ & DELPHI& 0 & \cite{delphi}\\
    \hline 
  \end{tabular} \end{center}
    \caption{\sf \label{tab:1} Experimental data used in the fits unless otherwise specified. }
\end{table}These observables can be expressed as  double expansions in $\alpha_s$ and 
inverse powers of $m_b$, schematically
\be
M_i= M_i^{(0)}+ \frac{\alpha_s(\mu)}{\pi} M_i^{(1)}+ \left(\frac{\alpha_s}{\pi}\right)^2 M_i^{(2)} + M_i^{(\pi)} \frac{\mu_\pi^2}{m_b^2}+ M_i^{(G)} \frac{\mu_G^2}{m_b^2}
+ M_i^{(D)} \frac{\rho_D^3}{m_b^3}+ M_i^{(LS)} \frac{\rho_{LS}^3}{m_b^3} +\dots
\label{double}
\ee
where all the coefficients $M_i^{(j)}$ depend on $m_c$, $m_b$,  
  $E_{cut}$, and on various renormalization scales. 
The OPE parameters $\mu_\pi^2$, ... are matrix elements of local $b$-fields operators 
evaluated in the physical $B$ meson, {\it i.e.}\ without taking the infinite mass limit. The dots represent 
missing terms of $O(\alpha_s/m_b^2)$ and $O(1/m_b^4)$, which are either 
 unknown or 
we do not include for the reasons  explained in the Introduction. We work in the 
kinetic scheme \cite{kinetic} and follow the implementation described in \cite{btoc,NNLO}. In 
particular, in the hadronic moments we do not expand in powers of $\bar\Lambda=M_B-m_b
$. While we always express the bottom mass and the four relevant expectation values that 
appear in Eq.~(\ref{double}) in the kinetic scheme setting the cutoff $\mu^{kin}$ at $1\GeV$, we will
use both  the kinetic  and the $\overline{\rm MS}$ scheme for the 
charm mass, and denote it with $m_c^{kin}(\mu^{kin})$ and $\overline{m}_c(\bar \mu)$, respectively. Unless otherwise specified we evaluate $\as$ at $\mu_0=4.6\GeV$ and assume $\as(\mu_0)=0.22$. A change of $\pm0.005$ around this value leads to  small changes in the results of our  fits (about $1 \rm MeV$ in $m_b$), always much smaller than their final uncertainty.

The experimental data for the moments are fitted to the theoretical expressions in order to gain 
information on the 
non-perturbative parameters and the heavy quark masses, which we then employ to extract
$|V_{cb}|$. 
Table \ref{tab:1} shows the 43  measurements of the moments which we always 
include in the fits, unless otherwise  
specified. The chromomagnetic expectation value $\mu_G^2$ is also constrained 
by the hyperfine splitting
\be 
M_{B^*}-M_B= \frac23 \frac{\mu_G^2}{m_b} + O\left(\frac{\as\mug}{m_b},\frac1{m_b^2}\right)\nonumber.
\ee 
Unfortunately, little is known of the power
corrections to the above relation and only a loose bound can be set, see \cite{long} for a recent discussion. For what concerns $\rho_{LS}^3$, it is somewhat constrained by the heavy quark sum rules. Following \cite{chromo,long,btoc}, we will therefore use in our fits the constraints 
\be
\mu_G^2=(0.35\pm 0.07)\GeV^2,
\qquad \rho_{LS}^3=(-0.15\pm 0.10 )\GeV^3\, . \label{constraints}
\ee
It should be stressed that $\rls$ plays a minor role in the fits because its coefficients are 
generally suppressed with respect to the other parameters.

We now perform a first global fit, without any theoretical uncertainty. The fit is not  good, 
 with   $\chi^2/dof\sim 2$, corresponding to a very small $p$-value and driven by a strong tension ($\sim 3.5 \sigma$) between the constraints in Eq.~\ref{constraints} and the measured moments. If we drop the constraints of Eq.~\ref{constraints} the fit is not too bad.
It is then clear from the outset that theoretical uncertainties are not so much necessary for the OPE expressions to fit the moments --- that would merely test Eq.(\ref{double}) as a parameterization;   they are instead needed to preserve the definition  of the parameters as $B$ expectation values of 
certain local operators,  which in turn can  be employed in the semileptonic widths and 
in other applications of the Heavy Quark Expansion.

\section{Theoretical errors and their correlations} 
The OPE description of semileptonic moments is subject to two sources of theoretical 
uncertainty: missing higher order terms in 
Eq.~(\ref{double}) and terms that violate  quark-hadron duality, namely terms that are 
intrinsically beyond the OPE.  We will attempt an estimate only of the first kind of uncertainty. 
The violation of local quark-hadron duality is expected to be suppressed in semileptonic $B$ 
decays; it would manifest itself as an inconsistency of the fit, which as we will see 
is certainly not present at the current level of theoretical and experimental accuracy.

To estimate the effects due to higher order corrections we follow the method outlined in 
\cite{btoc} and update it with the suggestions given in \cite{NNLO}: we assume that perturbative corrections can affect
 the Wilson coefficients of $\mu_\pi^2$ 
and $\mu_G^2$  at the level of $\pm 20\%$, while perturbative 
corrections and higher power corrections can effectively change the coefficients of $
\rho_D^3$ and $\rho_{LS}^3$ by $\pm 30\%$. Moreover we assign an irreducible theoretical 
uncertainty of 10 MeV to the heavy quark masses, and vary $\alpha_s(m_b)$ by 0.02. 
The changes in $M_i$ due these variations of the fundamental parameters are added in quadrature and provide a theoretical uncertainty\footnote{This theoretical uncertainty depends of course on the exact point in the parameter space. In practice we adopt an iterative procedure recomputing the theory errors on the minimum $\chi^2$ point till the process has converged.}
 $\delta M_i^{th}$, to be subsequently added 
in quadrature with the experimental one, $\delta M_i^{exp}$.
This method is consistent with the residual scale dependence observed at NNLO, and appears to be reliable: the NNLO corrections and the $O(1/m_b^{4,5})$ (using 
ground state saturation as in \cite{Mannel:2010wj}) have been found to be within the range 
of expectations based on the method in the original formulation of \cite{btoc}. 

The correlation between  theoretical errors assigned to different observables are much 
harder to estimate, but they play an important role in the semileptonic fits, as it will become 
clear in a moment. Let us first consider moments computed at a fixed value of $E_{cut}$: as 
long as one deals with central higher moments, there is no argument of principle supporting a 
correlation between two different moments, for instance $\ell_1$ and $h_2$. We also do not observe 
any clear pattern in the known corrections, and therefore regard the theoretical predictions 
for  different central moments as completely uncorrelated. When one computes the theory 
uncertainty with the method described above, one might think there are obvious correlations: 
suppose that both $\ell_1$ and $h_2$ receive {\it positive} contributions from $\mu_\pi^2$; 
by varying  $\mu_\pi^2$ by $\pm20\%$ we see a {\it positive} correlation between $\ell_1$ and 
$h_2$.
But our aim was simply to have a rough estimate of the {\it size} of the theory uncertainty, and we make no 
claim to know  the {\it exact magnitude} and {\it sign} of the higher-order correction! Therefore, the observed correlation is meaningless and the safest assumption is 
 to regard the theoretical predictions for different moments as uncorrelated.

Let us now consider the calculation of a certain moment $M_i$ for two close values of $E_{cut}$, say 1\GeV\ and 1.1\GeV. Clearly, the OPE expansion for $M_i(1\GeV)$ will
be very similar to the one for $M_i(1.1\GeV)$, and we confidently expect this to be true 
at any order in $\as$ and $1/m_b$. The theoretical uncertainties we assign to $M_i(1\GeV)$
and $M_i(1.1\GeV)$ will therefore be very close to each other and {\it very highly 
correlated}. The degree of correlation between the theory uncertainty of $M_i(E_1)$ and 
$M_i(E_2)$ can intuitively be expected to decrease as $|E_1-E_2|$ grows.
Moreover, we know that higher power corrections are going to modify significantly the 
spectrum only close to the endpoint. Indeed, one observes  that the $O(1/m_b^{4,5})$ 
contributions are equal for all cuts below about 1.2\GeV\ (see Fig.2 of \cite{Mannel:
2010wj}) and the same happens for the $O(\as\mupi/m_b^2)$ corrections \cite{Becher:
2007tk}. Therefore, the dominant sources of current theoretical uncertainty
suggest very high correlations among the theoretical predictions of the moments for cuts
below roughly 1.2 \GeV.

 In Refs.~\cite{babarfit,BF},  it has been assumed that the theoretical errors of moments  
 at different 
values of $E_{cut}$ are 100\% correlated. This is too strong an assumption, which ends up distorting the fit, 
because the dependence of $M_i$ on $E_{cut}$, itself a function of 
the fit parameters, is  then free of theoretical uncertainty. 
As a result, the uncertainty on the OPE parameters is underestimated.
On the other hand, as we have just discussed, a high degree of correlation is to be 
expected if $E_{cut}$  is not too large.
 Ref.~\cite{1S}, which also presents a fit without theoretical errors, sets the theoretical correlation matrix equal to the experimental one. This 
 probably underestimates correlations, and  would also imply no correlation between the 
 theoretical prediction of the {\em same} observable measured by two independent experiments, which is unreasonable. Recent HFAG fits with the method of  Ref.~\cite{1S}
do not present this problem.
 
 An alternative approach consists in computing the correlation of 
the theoretical errors of $M_i(E_1)$ and $M_i(E_{2})$ using the method
we have used to estimate the theory uncertainty, namely varying the values of the HQE parameters.   This is in fact equivalent to 
assuming that the $E_{cut}$-dependence of the unknown terms of the OPE follows 
closely the $E_{cut}$-dependence of the known terms. It turns out that in this 
way the correlation is almost always very high.

Another possibility is to fix the correlation $\xi$ between a moment $M_i$ computed at  
$E_{cut}$ and at $E_{cut}+0.1\GeV$, possibly with $\xi$ higher for higher $E_{cut}$.
Following the reasoning above, $\xi$
is naturally very close to 1 at low cuts, and drops considerably at high cuts.

In our fits, we will consider the following four options:
\begin{itemize}
\item [{\bf A}] 100\% correlation between moments at different cuts; 
\item [{\bf B}] correlations computed from theory predictions, as discussed above;
\item [{\bf C}] constant scale factor $0<\xi<1$, with $\xi=0.97$ for 100 MeV steps;
\item [{\bf D}] a scale factor like in {\bf C} that depends on the cut,  $\xi=\xi(E_{cut})$, with
\be 
\xi(E_{cut})= 1- \frac12 \,e^{-\frac{(E_0-E_{cut})}{\Delta}}
\ee
where $E_0\approx 1.75\GeV$ is the partonic endpoint and $\Delta$ is
an adjustable parameter which we set at about $0.25\GeV$.
\end{itemize}
As an illustration,
 the correlation between the theoretical errors of a generic moment with cuts at 0.6 and 1.2 GeV is 1  in scenario {\bf A}; it depends on the specific moment and it is generally quite close to 1 
 in scenario {\bf B};  it is given by  $\xi^6=(0.97)^6\simeq 0.83$ in scenario {\bf C}; it is given by 
$\prod_{k=0}^{k=5} \xi(0.65+ 0.1 k\GeV)\simeq 0.88$ in scenario {\bf D}.
Similarly, the correlation between a moment measured at 
1.3 and 1.5\GeV\ is approximately 0.94 in {\bf C} and 0.76 in {\bf D}.

\begin{figure}
\begin{center}
\includegraphics[width=6.9cm]{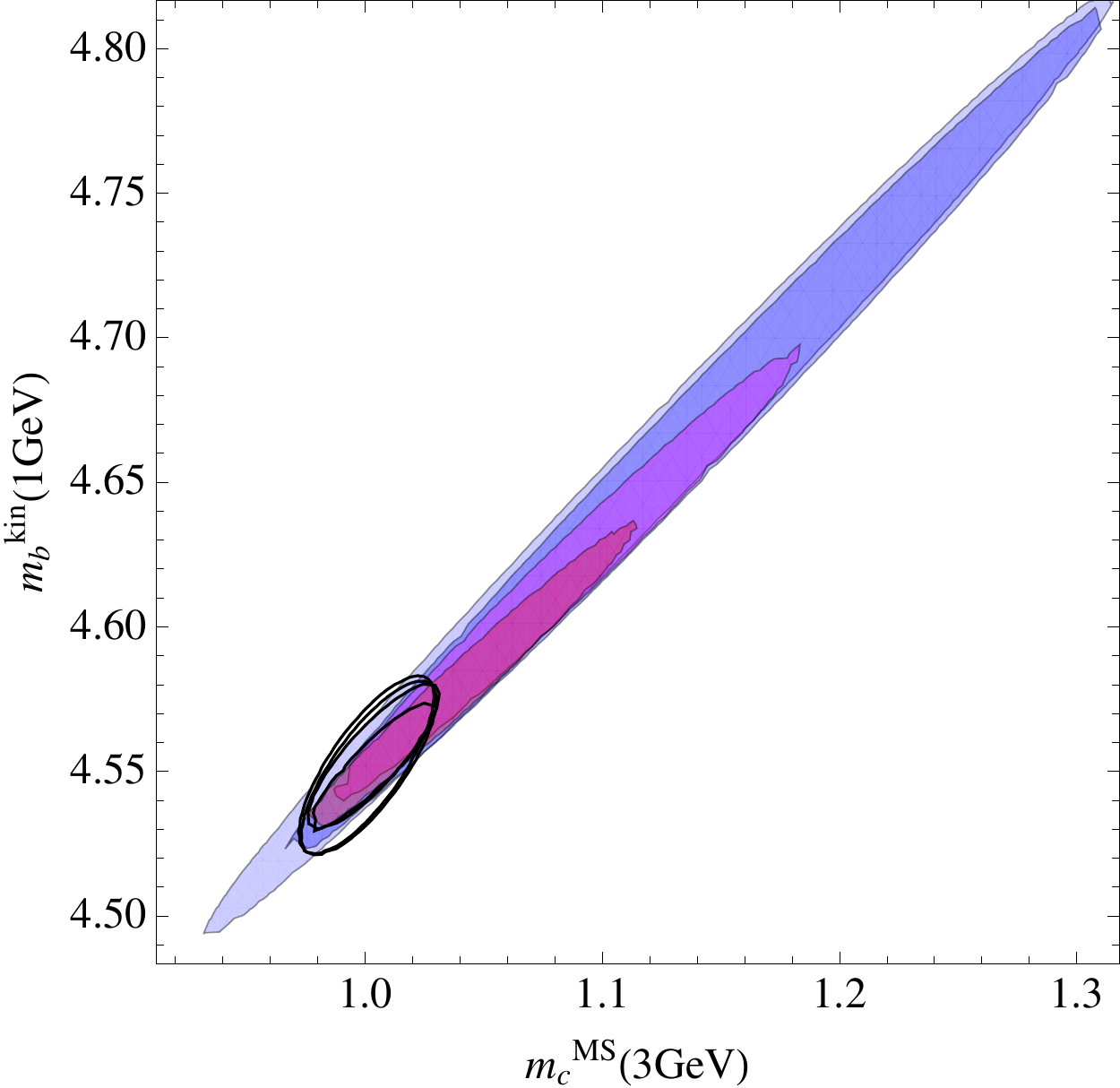}
\includegraphics[width=6.9cm]{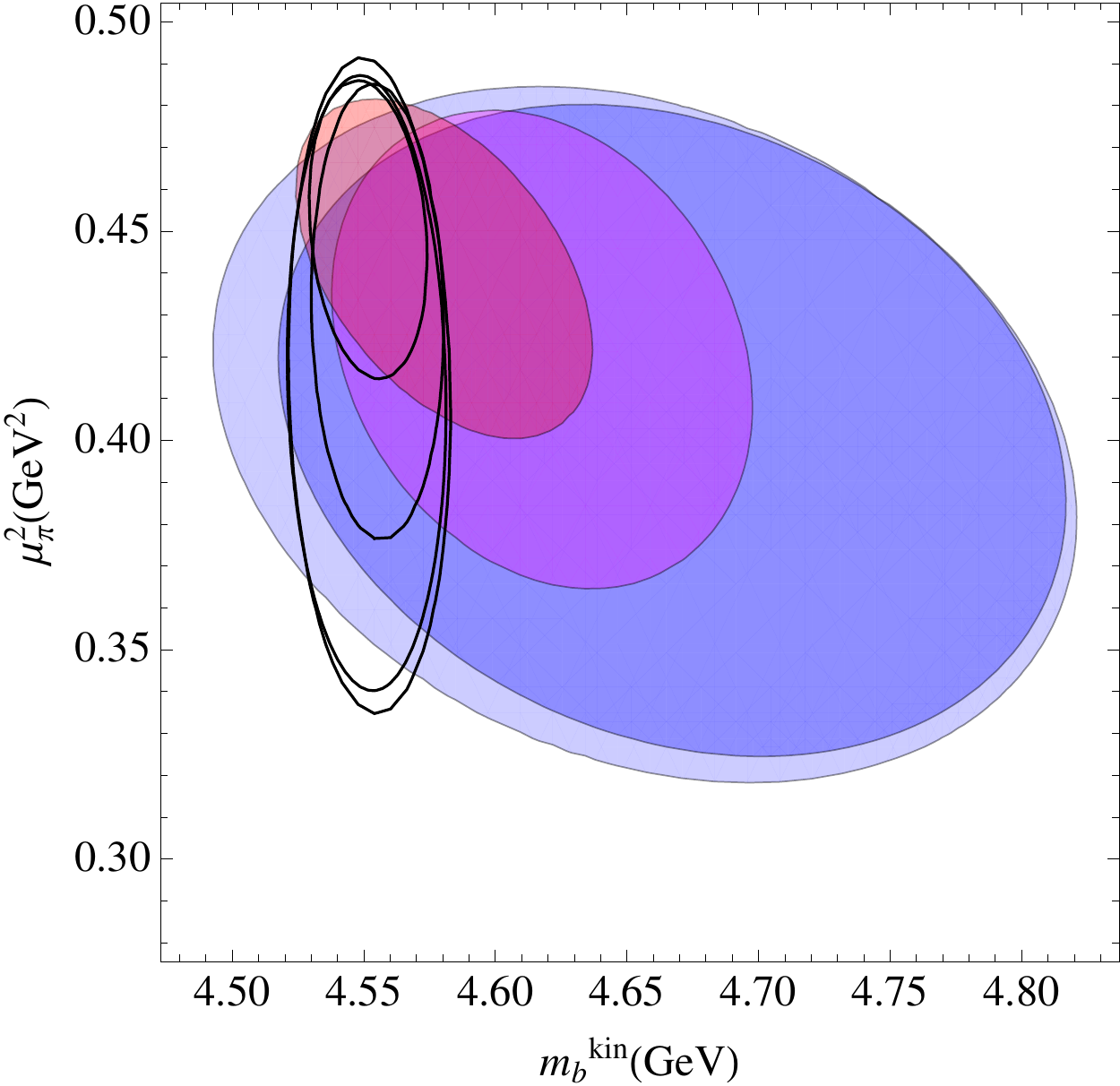}\\
\includegraphics[width=6.9cm]{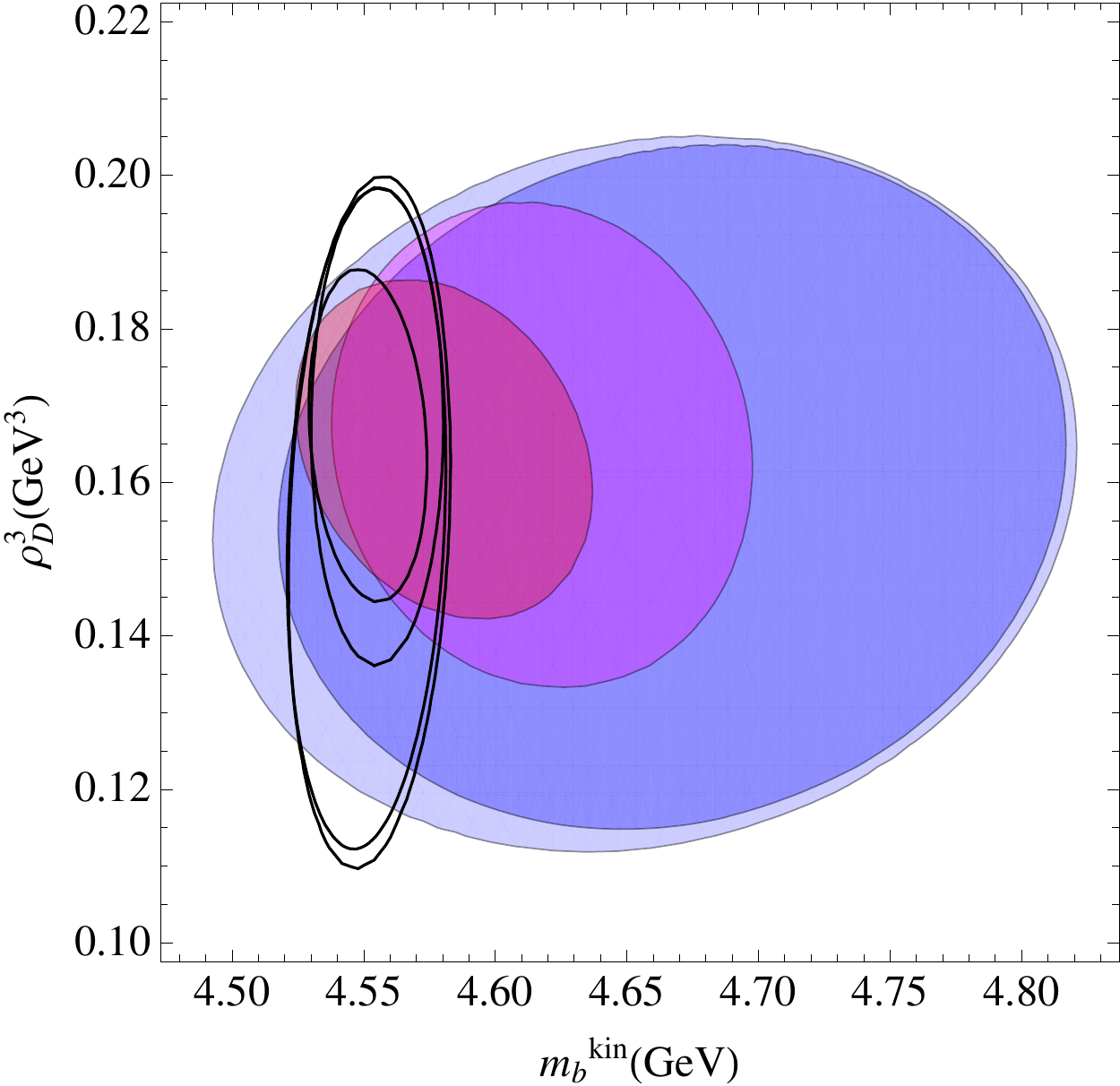}
\includegraphics[width=6.9cm]{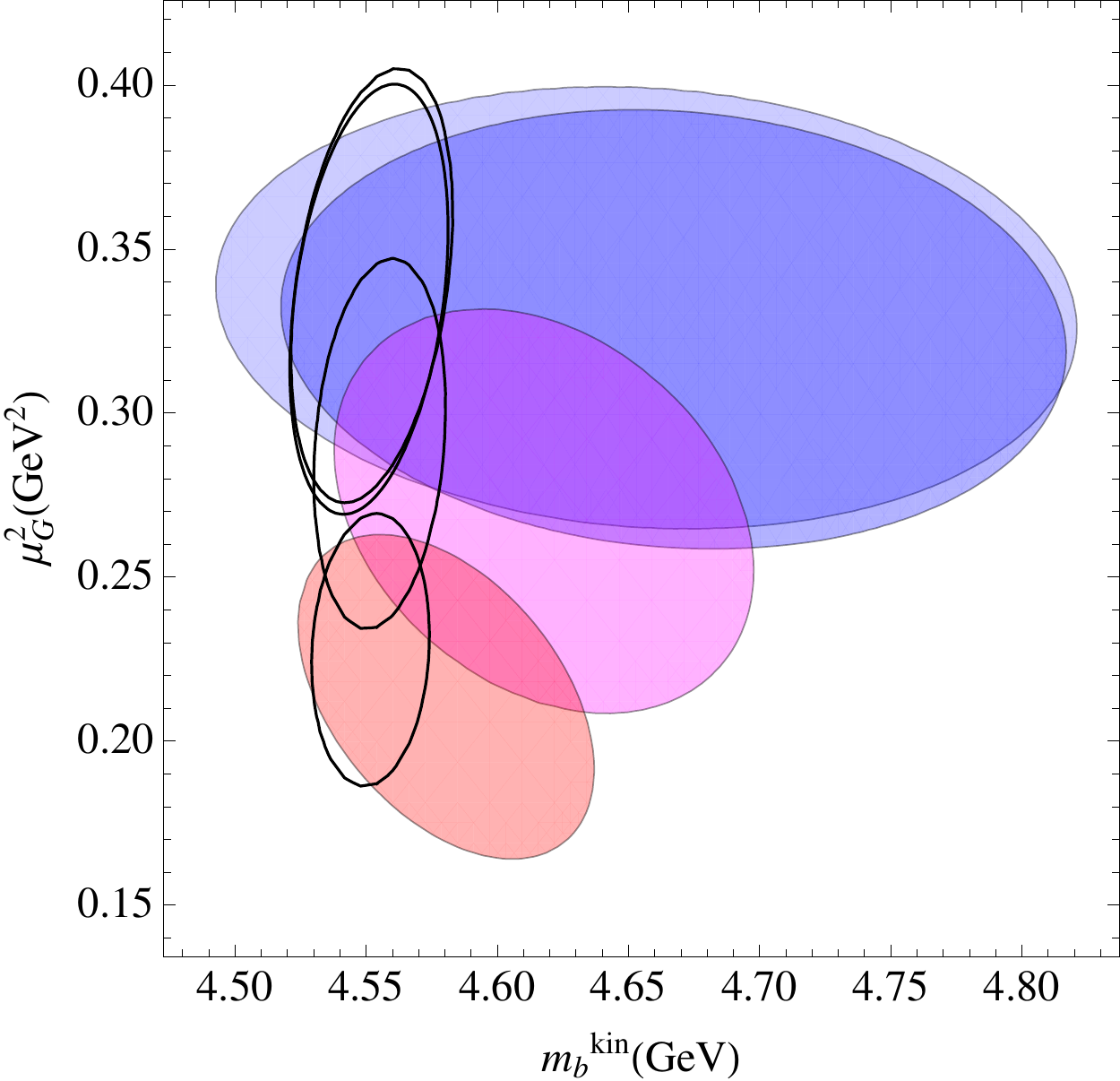}\\
\includegraphics[width=6.9cm]{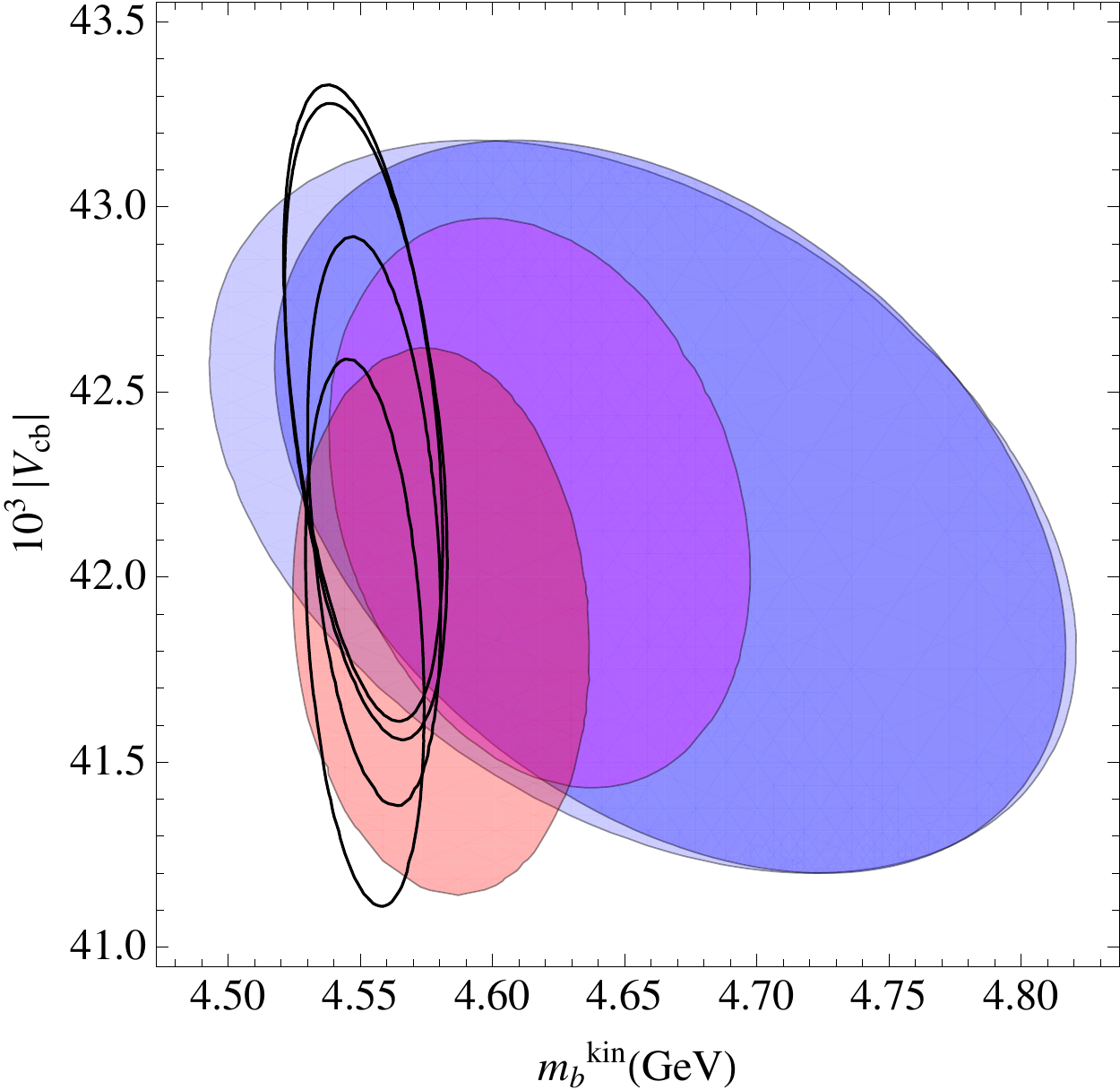}
\includegraphics[width=6.9cm]{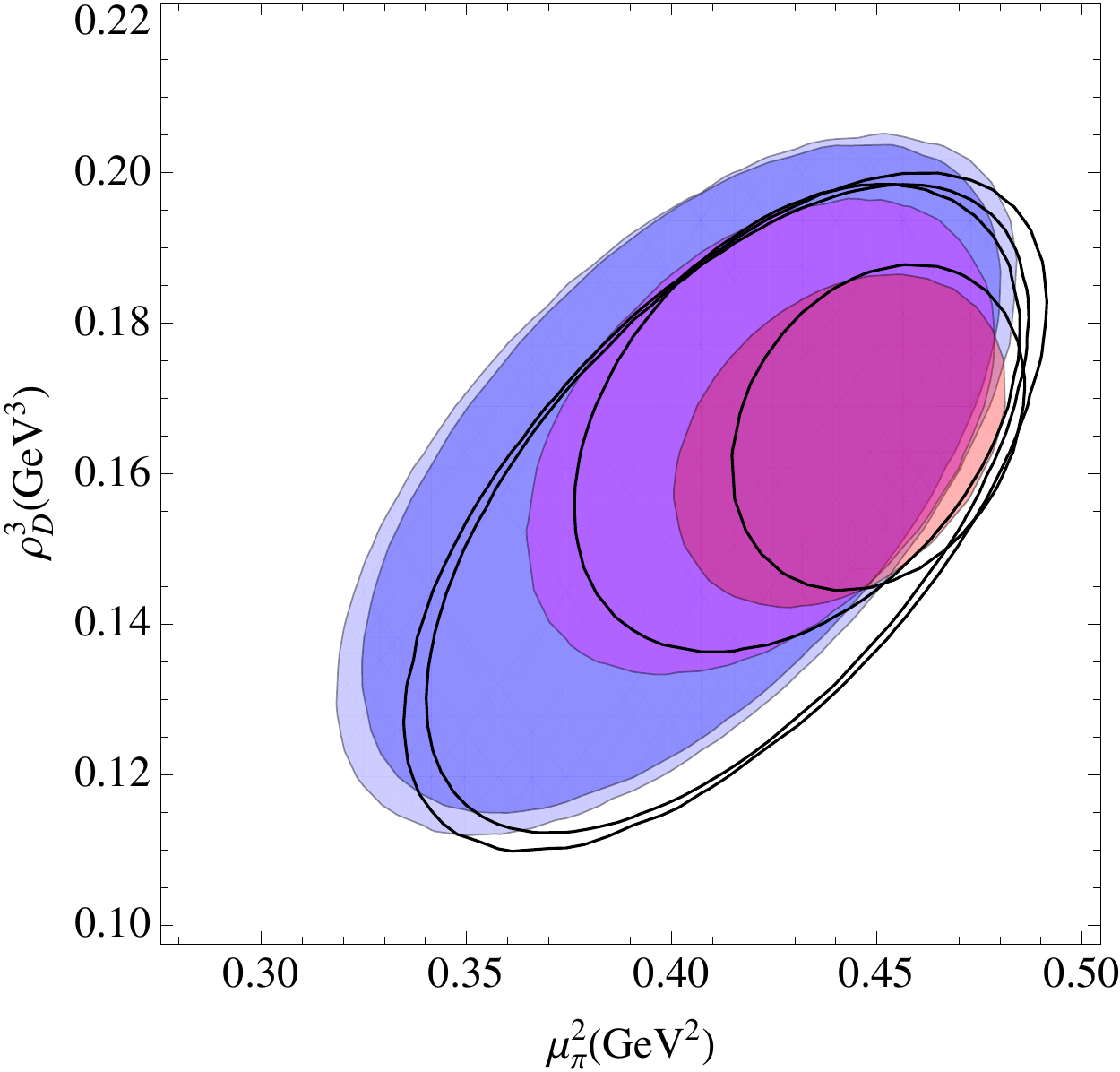}
\caption{\sf Two-dimensional projections of the fits performed with different assumptions for the theoretical correlations. The orange, magenta, blue, light blue 1-sigma regions correspond to scenarios A,B,C,D ($\Delta=0.25$GeV), respectively. The black contours show  the same regions when the $m_c$ constraint of Ref.~\cite{mc_hoang} is employed.
}
\label{fig1}
\end{center}
\end{figure}

In Fig.~1 we show the results of semileptonic fits performed with the four options for the 
 theoretical correlations. The fits include all the data listed in Table 1 and the two 
 constraints of Eq.~(\ref{constraints}). We add the experimental and theoretical covariance matrices, which is equivalent to adding the respective errors in quadrature.
 
In general, the results depend sensitively on the option adopted. In the case of the heavy 
quark masses, which are strongly correlated, we observe  large errors that tend to increase in going from {\bf A} to {\bf D}, although the central values are quite consistent.  
The results of the fits for the non-perturbative parameters depend even stronger on the 
option, in particular this is the case for $\mu_G^2$, which has a very low value --- incompatible with (\ref{constraints}) --- in  scenario {\bf A}, and increases as correlations are relaxed.

Fig.~1 also shows that the final uncertainty on some of the parameters can be much smaller 
than the "safety" range we have used in the evaluation of the theory errors. Consider for 
instance the final error on $\rd$: in scenario {\bf A} it is as low as 0.02, which is 
approximately 15\% of the central value, much below the 30\% we have employed.
Evidently, the fit has found a direction in the parameter space with much lower uncertainty.
On the other hand, when we relax the degree of correlation, as in options {\bf C} and {\bf D},
the final relative uncertainty on $\rd$ is close to 30\%. 
In the next Section we will study the impact of additional constraints on the heavy quark 
masses on these fits.

\section{Heavy quark mass constraints}
The inclusion of external precise constraints
in the fit decreases the errors and may neutralize the  ambiguity
due to the ansatz for the theoretical correlations. It also allows us to check the consistency 
of the results with independent information.
 As semileptonic $B$ decays alone determine precisely a linear combination 
of the heavy quark masses, approximately given by $m_b-0.8\,m_c$, see the first plot in Fig.~1, a way to maximally exploit their potential consists in including in the fit one of the recent 
precise $m_c$ determinations.
A review of  heavy quark mass determinations is beyond the scope of this paper, see {\it e.g.}\ \cite{Antonelli:2009ws,PDG}. We simply list some of  the most recent ones, for the charm mass 
\begin{enumerate}
\item $\overline m_c(3\GeV)=0.986(13)$GeV \cite{mcmb_karlsruhe};
\item $\overline m_c(3\GeV)=0.986(6)$GeV \,\, \cite{masseslat};
\item $\overline m_c(3\GeV)=0.994(26)$GeV \cite{mc_hoang};
\end{enumerate}
and for the bottom mass
\begin{enumerate}
\item $\overline m_b(\overline m_b)=4.163(16)$GeV \cite{mcmb_karlsruhe};
\item $\overline m_b(\overline m_b)=4.164(23)$GeV \cite{masseslat};
\item $\overline m_b(\overline m_b)=4.235(55)$GeV \cite{mb_hoang};
\item  $\overline m_b(\overline m_b)=4.247(34)$GeV \cite{simula}.
\end{enumerate}
Here all the masses  are expressed in  the $\overline{\rm MS}$ scheme, and 
a relatively high scale, 3\GeV, is employed for the charm mass. 
In absolute terms, the charm mass is currently better determined than the bottom mass.  This suggests to compute the moments 
directly in terms of $\overline m_c(\bar\mu)$, with $2\lsim \bar \mu\lsim 3\GeV$, instead of using 
the charm mass in the kinetic scheme.  The range of $\bar\mu$ is chosen to
avoid large logarithms in our $O(\as^2)$ calculation and to minimize higher orders related 
to the definition of $m_c$, which necessarily involve $\as(\bar\mu)$.
This  $\overline{\rm MS}$
option for $m_c$ is available in the code of \cite{NNLO} and avoids  additional 
theoretical uncertainty due to the mass scheme conversion. In the case of the bottom mass,
on the contrary, the common choice  $\overline m_b(\overline m_b)$ is not  well-suited to the description of semileptonic $B$ decays. In other words, the calculation of the moments in terms of $\overline m_b(\overline m_b)$ would lead to large 
higher order corrections. While
our predictions are always expressed  in terms of the kinetic mass $m_b^{kin}(1\GeV)$,
the above $m_b$ constraints can be included after converting them to the kinetic scheme.
Since the relation between the kinetic and the $\overline{\rm MS}$  masses is known only to 
$O(\as^2)$, the ensuing uncertainty is not negligible. In Ref.~\cite{NNLO} it has been estimated to be about 30 MeV: 
\be\label{transl}
m_b^{kin}(1\GeV)-\overline m_b(\overline m_b) = 0.37\pm 0.03 \GeV.
\ee

The effect of the inclusion of charm mass constraints in the semileptonic fit is illustrated in 
Fig.~1, where the determination of Ref.~\cite{mc_hoang} is employed. As expected, the 
uncertainty in the $b$ mass becomes smaller than 30MeV in all scenarios, a marked improvement, also with respect to the precision resulting from the use of radiative moments \cite{HFAG}. On the other 
hand, there is hardly any  improvement  in the final precision of the non-perturbative 
parameters (see for instance the last plot, in the $\mupi$-$\rd$ plane). 

As already noted,  the semileptonic moments are highly sensitive to a linear 
combination of the heavy quark masses. The constraints on $m_b$ that we obtain 
using different $m_c$ determinations in the fit are shown in Table 2, where we have only 
considered option {\bf D} with $\Delta=0.25\GeV$.
\begin{table}[t]
  \begin{center} \begin{tabular}{|c|c|c|}
    \hline 
  $\overline m_c(3\GeV)$   & $m_b^{kin}(1\GeV)$ & $\overline m_b(\overline m_b) $\\ \hline \hline
    0.986(13) \cite{ mcmb_karlsruhe} & 4.541(23) &  4.171(38) \\  \hline
      0.986(6) \ \cite{ masseslat} & 4.540(20) &  4.170(36)   \\  \hline
     0.994(26) \cite{mc_hoang} &  4.549(29) & 4.179(42) \\ \hline
  \end{tabular} \end{center}
    \caption{\sf \label{tab:2} $ b$ mass resulting from different $m_c$ determinations.  All masses are expressed in\,\GeV. }
\end{table}
It is interesting to compare the results in the last column with the $\overline m_b(\overline 
m_b)$ determinations we have listed above. The bottom mass obtained using  $m_c$
given by the Karlsruhe group \cite{mcmb_karlsruhe} is perfectly consistent with their own
$m_b$ result, but also compatible with those of Refs.~\cite{mb_hoang,masseslat}.
In general, lower values of $m_b$ are preferred.
The results depend little on the scenario chosen for the theory correlations: if we choose $\Delta=0.2\GeV$
 $m_b$  gets lowered by 1 MeV in the first two rows, and by 2 MeV in the third. Very similar 
 results are found using alternative scenarios for the theory correlations. Of course, one can 
 also include in the fit both $m_c$ and $m_b$ determinations, 
but because of the scheme translation error in $m_b$ the gain in accuracy will be  limited.

When no external constraint is imposed on $m_{c,b}$, the semileptonic moments determine best   a linear combination of the heavy quark masses which is very close to their difference. Using scenario {\bf D} with $\Delta=0.25\GeV$ we obtain
\be
m_b^{kin}(1\GeV)-  0.85\, \overline m_c(3\GeV)= 3.701\pm 0.019\GeV, \label{massdiff}
\ee
and  similar results with the other scenarios  (the error is as low as 12 MeV in scenario {\bf A}). The ratio of the two masses is $\overline m_c(3\GeV)/m_b^{kin}(1\GeV)=0.2172(25)$.
In the case the kinetic scheme is also adopted for $m_c$, the linear combination is slightly different and Eq.~(\ref{massdiff}) becomes
 \be
m_b^{kin}(1\GeV)-  0.7\, m_c^{kin}(1\GeV) = 3.784 \pm 0.019\GeV. 
\ee

\begin{table}[t]
  \begin{center} \begin{tabular}{|c|ccccccc|c|}
    \hline 
{\footnotesize th.\,corr.\,scenario} &$m_b^{kin}$ & $ m_c$   &  $\mupi $ &$\rd$ &$\mug$ & $\rls$  & ${\rm BR}_{c\ell\nu}${\footnotesize (\%)}& $10^3 \,|V_{cb}|$ \\ \hline\hline
   {\bf D}  \cite{mcmb_karlsruhe} & 4.541 &0.987 & 0.414 & 0.154 & 0.340 & -0.147 & 10.65 & 42.42 \\  
   {\footnotesize $\overline{m}_c(3{\rm GeV})$} & 0.023 & 0.013 & 0.078 & 0.045 & 0.066 & 0.098 & \ 0.16 & \ 0.86
   \\  \hline   
        {\bf A} \cite{mcmb_karlsruhe} & 4.540 &0.987 & 0.454 & 0.167 & 0.234 & -0.078 & 10.45 & 41.85 \\  
    {\footnotesize $\overline{m}_c(3{\rm GeV})$} & 0.014& 0.013 & 0.035 & 0.022 & 0.040 & 0.085 & \ 0.13 & \ 0.74  \\  \hline
            {\bf B} \cite{mcmb_karlsruhe} & 4.542 &0.987 & 0.457 & 0.184 & 0.290 & -0.135 & 10.51 & 42.15 \\  
  {\footnotesize $\overline{m}_c(3{\rm GeV})$}   & 0.017& 0.013 & 0.056 & 0.035 & 0.056 & 0.095 & \ 0.14 & \ 0.77  \\  \hline
     {\bf C} \cite{mcmb_karlsruhe} & 4.539 &0.987 & 0.415 & 0.155 & 0.336 & -0.147 & 10.65 & 42.45 \\  
   {\footnotesize $\overline{m}_c(3{\rm GeV})$}  & 0.022 & 0.013 & 0.073 & 0.043 & 0.066 & 0.098 & \ 0.16 & \ 0.86  \\  \hline
           {\bf D} \cite{mcmb_karlsruhe} & 4.538 &0.986 & 0.415 & 0.153 & 0.336 & -0.145 & 10.65 & 42.46 \\  
 {\footnotesize $\overline{m}_c(3{\rm GeV}), m_b$}   & 0.018 & 0.012 & 0.078 & 0.045 & 0.064 & 0.098 & \ 0.16 & \ 0.84
   \\  \hline
    {\bf D}  \cite{mc_hoang} & 4.549 &0.996 & 0.413 & 0.154 & 0.339 & -0.146 & 10.65 & 42.40 \\  
 {\footnotesize $\overline{m}_c(3{\rm GeV})$}   & 0.029 & 0.026 & 0.078 & 0.045 & 0.066 & 0.098 & \ 0.16 & \ 0.87
   \\  \hline
      {\bf D}  \cite{mcmb_karlsruhe} & 4.548 &1.092 & 0.428 & 0.158 & 0.344 & -0.146 & 10.66 & 42.24 \\  $m_c^{kin}$
   & 0.023 & 0.020 & 0.079 & 0.045 & 0.066 & 0.098 & \ 0.16 & \ 0.85
   \\  \hline
         {\bf D}  \cite{mcmb_karlsruhe} & 4.553 &1.088 & 0.428 & 0.155 & 0.328 & -0.139 & 10.67 & 42.42 \\ {\footnotesize $\overline{m}_c(2{\rm GeV}),m_b$}
   & 0.018 & 0.013 & 0.079 & 0.045 & 0.064 & 0.098 & \ 0.16 & \ 0.83
   \\  \hline
   \end{tabular} \end{center}
    \caption{\sf \label{tab:3} Global fits  with $m_c$ constraints. Scenario {\bf D} has $\Delta=0.25\GeV$. All parameters except  $m_c$ are in the kinetic scheme with cutoff at 1\GeV. The definition of $m_c$ and the use of an $m_b$ constraint are marked  in the first column, directly under the reference for their constraints.}
\end{table}
\begin{table}[t]
  \begin{center} \begin{tabular}{|cccccccc|}
    \hline 
 $m_b^{kin}$ & $\overline m_c(3\GeV)$   &  $\mupi $ &$\rd$ &$\mug$ & $\rls$  & ${\rm BR}_{c\ell\nu}$ & $|V_{cb}|$ \\ \hline
   1 & 0.476 & -0.101 & 0.218 & 0.484 & -0.158 & -0.092 & -0.443\\
  & 1 & -0.013 & 0.009 & -0.014 & 0.004 & 0.012& -0.014\\
 &  & 1& 0.613 & 0.007 & 0.056 & 0.126 &0.342\\
  &  &  & 1 & -0.041 & -0.126 & 0.048& 0.179\\
  &  &  &  & 1 & -0.013 & -0.023& -0.164\\
  &  &  &  &  & 1 & -0.007 &0.009\\
  &  &  &  &  &  & 1 &    0.461\\
   &  &  &  &  &  & & 1    \\  \hline
   \end{tabular} \end{center}
    \caption{\sf \label{tab:4} Correlation matrix for the default fit: scenario {\bf D}, $\Delta=0.25\GeV$, $m_c$ from \cite{mcmb_karlsruhe}. }
\end{table}
\begin{table}
  \begin{center} \begin{tabular}{|cccccccc|}
    \hline 
 $m_b^{kin}$ & $\overline m_c(3\GeV)$   &  $\mupi $ &$\rd$ &$\mug$ & $\rls$  & ${\rm BR}_{c\ell\nu}$ & $|V_{cb}|$ \\ \hline
  1 & 0.405 & -0.082 & 0.180 & 0.412 & -0.130 & -0.075 &-0.389 \\
  & 1 & 0.003 & -0.027 & -0.098 & 0.030 & 0.028 &  0.041\\
  &  & 1& 0.626 & 0.025 & 0.051 & 0.123 & 0.338\\
  &  &  & 1 & -0.080 & -0.116 & 0.055 & 0.207\\
  &  &  &  & 1 & 0.013 & -0.008 & -0.120\\
  &  &  &  &  & 1 & -0.012 & -0.007\\
  &  &  &  &  &  & 1 & 0.463\\
  &  &  &  &  &  & & 1 
   \\  \hline
   \end{tabular} \end{center}
    \caption{\sf \label{tab:5} Correlation matrix for the fit with both  $m_c$ and $m_b$ from \cite{mcmb_karlsruhe}, 
         scenario {\bf D}, $\Delta=0.25\GeV$. }
\end{table}

The  results of a few  fits are reported in Table \ref{tab:3}.  We choose the first one as our {\it default} fit. All the fits  include a constraint on $m_c$, from either Ref.~\cite{mcmb_karlsruhe} or \cite{mc_hoang}, and  two fits  both mass constraints from Ref.~\cite{mcmb_karlsruhe}. In the latter case 
we have  used (\ref{transl}) to  translate $\overline m_b(\overline m_b)=4.163(16)\GeV$ into $m_b^{kin}=4.533(32)\GeV$ (the $\as$ dependence of Eq.~(\ref{transl}) partly compensates that of $\overline m_b(\overline m_b)$).
The fits are generally  good, with  $\chi^2/d.o.f.$ ranging from 0.32  for the default fit, to 0.95 for case {\bf B} and   1.18 for  case {\bf A}.
The value of $|V_{cb}|$ is computed using
\be
|V_{cb}|=\sqrt{\frac{|V_{cb}|^2\, {\rm BR}_{c\ell \nu}}{\tau_B \,\Gamma_{B\to X_c \ell \nu}^{OPE}} ,
}\ee
with $\tau_B= 1.582(7)$\,ps. Its theoretical error is computed combining in quadrature the parametric uncertainty that results from the fit, and an additional 1.4\% theoretical error
to take into account missing higher order corrections in the expression for the semileptonic 
width \cite{imprecated,NNLO}. An approximate formula for $|V_{cb}|$ using the above $\tau_B$ value and $m_c(3\GeV) $ is
\bea
|V_{cb}|\!&=& \!\!\!0.042316 \Big[1+0.54\,(\as\!-\!0.219)-0.653\, (m_b^{kin}\!-\!4.55)+0.489\,(\overline{m}_c(3\GeV)\!-\!1) \label{unfactor}
\nonumber\\&&\!\!\!+0.016 \,(\mu_\pi^2\!-\!0.44)+0.058\,(\mu_G^2\!-\!0.32)+0.12\, (\rho_D^3\!-\!0.2)-0.013\, (\rho_{LS}^3\!+\!0.15)\Big],
\eea
where all dimensionful quantities are expressed in GeV.

A few comments are now  in order:
\begin{itemize}
\item [\it i)] The inclusion of the $m_c$ constraint has stabilized the fits with respect to the ansatz for the theory correlations. The only exception is represented by scenario {\bf A}, which mostly deviates in the values of $\mupi$ and $\mug$ and in the magnitude of the uncertainties. In any case, because of the above discussion, this scenario should be abandoned.
\item [\it ii)] The low $\chi^2$ of the default fit is due  to the large theoretical uncertainties we 
have assumed.
It may be tempting to interpret it as evidence that the theoretical errors 
have been overestimated. However,  higher order corrections may 
effectively shift the parameters of the $O(1/m_b^2)$ and  $O(1/m_b^3)$ contributions. If we 
want to maintain the formal definition of these parameters, and to be able to use them 
elsewhere, we therefore have to take into account the potential shift they may experience 
because of higher order effects. 

\item [\it iii)]  The third hadronic moment by Delphi was neglected in previous analyses in the kinetic scheme. Its main effect on our default fit is to decrease $\rd$ by about 10\% and $\mupi$ by about 3\%.  

\begin{figure}
\begin{center}
\includegraphics[width=6.9cm]{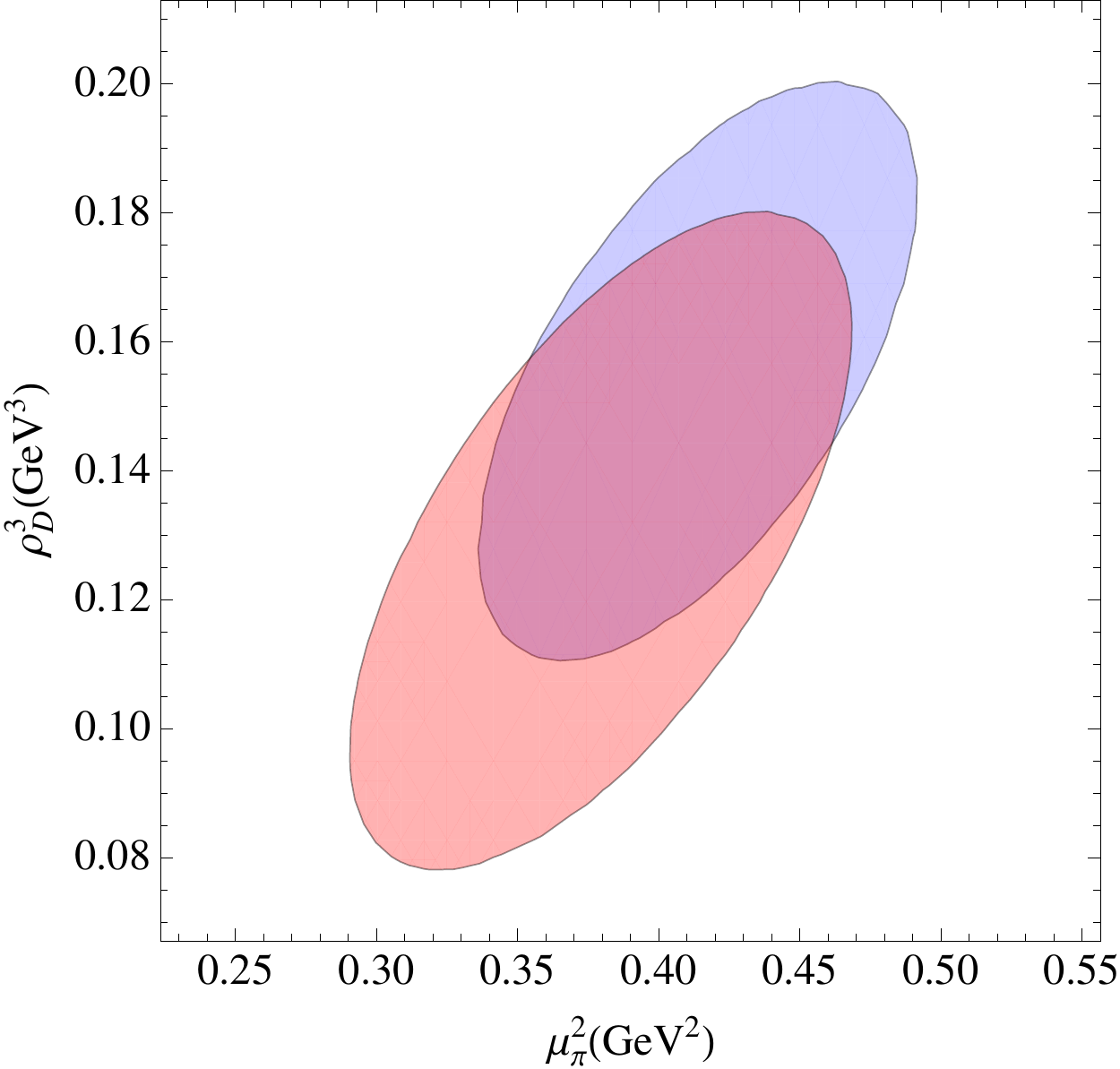}\ \ \
\includegraphics[width=6.9cm]{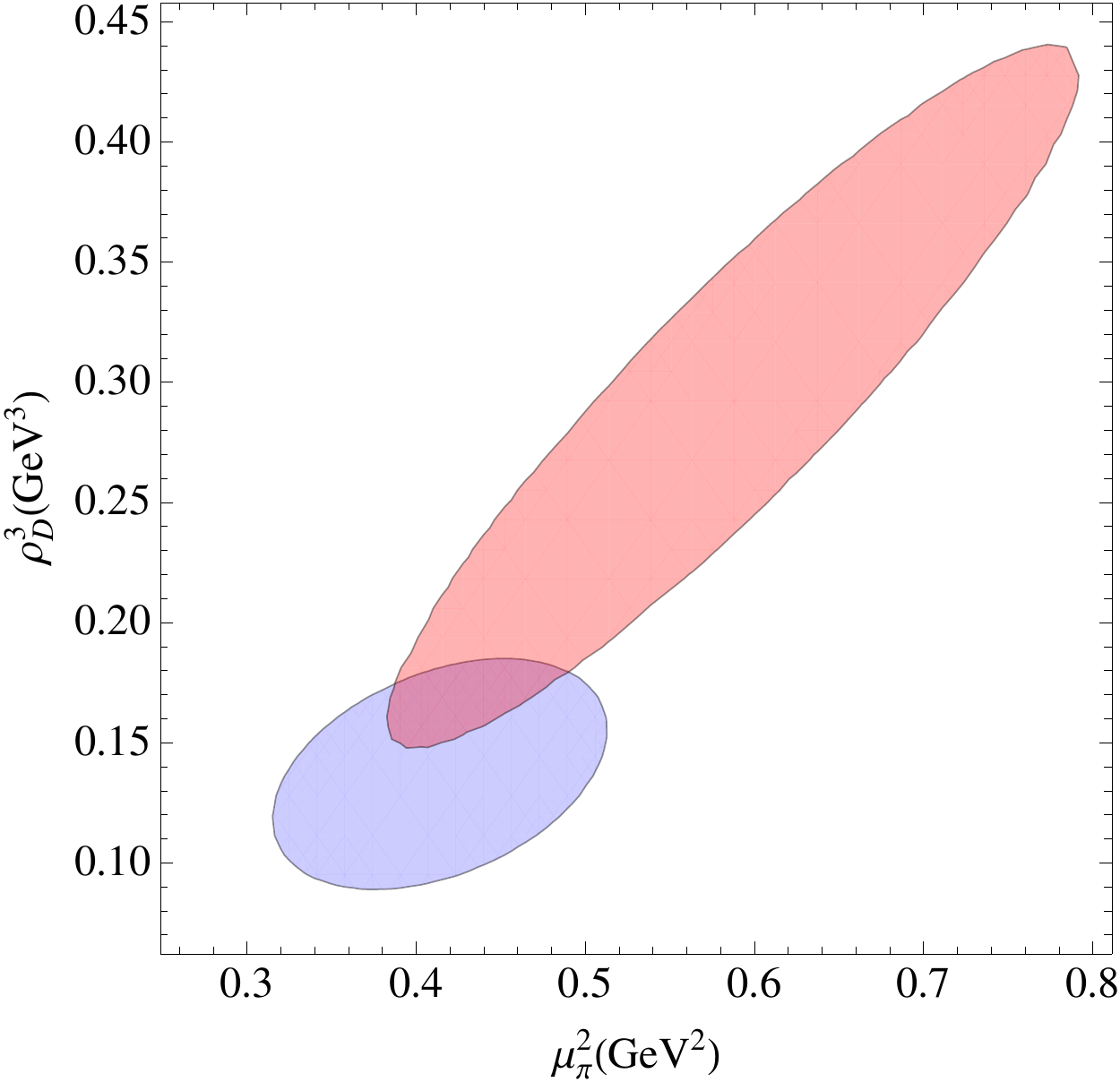}
\caption{\sf 1$\sigma$ projections on the $\mupi,\rd$ plane of the default fit. Left:  with (blue) and without (red)
moments measured at $E_{cut}>1.2\GeV$; right:  with only hadronic moments (blue) and only leptonic moments (red).}
\label{highcuts}
\end{center}
\end{figure}

\item [\it iv)] The fits with a constraint on $m_c$ are quite stable with respect to a  change of inputs. In particular, we 
have found small differences when experimental data at high $E_{cut}$ are excluded,
and when  only hadronic or leptonic moments are considered.  While the results for the heavy quark masses do not change appreciably, the results for the OPE parameters 
change within errors. We show in Fig.~\ref{highcuts} the projection onto the $\mupi-\rd$ plane of the default fit with/without moments at $E_{cut}>1.2\GeV$ and with only leptonic and only hadronic moments.

\item [\it v)] One may wonder whether the inclusion of moments measured at
different values of $E_{cut}$ really benefits the final accuracy. We have run
our default fit with only one $E_{cut}$ for each moment per experiment and found 
slightly larger errors (26\,MeV on $m_b$ and $0.1\GeV^2$ on $\mupi$) than when we use the full set of data.  The benefit is therefore minor but non-negligible. Of course, 
the inclusion of moments at different cuts plays a much more important role 
 in scenarios {\bf A} and {\bf B}.
 
\item [\it vi)] In the kinetic scheme the inequalities
$\mu_\pi^2(\mu)\!\ge\!\mu_G^2(\mu)$, $\rd(\mu)\!\ge\!-\rls(\mu)$
hold at arbitrary values of the cutoff $\mu$. The central values of the fits always
satisfy the inequalities. 

\item [\it vii)] The value of $|V_{cb}|$ is generally larger than in previous analyses. This is 
mostly due to the higher BR$_{c\ell \nu}$, as can be seen in Table 3. Indeed, the low value of BR$_{c\ell \nu}$ is the most distinctive feature of scenario {\bf A} and {\bf B}, and the most relevant for the $|V_{cb}|$ determination.
  It is worth noting that  the latest CKM global fit  \cite{UTfit} gives  $|V_{cb}|=0.04273(77)$, with a marked preference for a high value of  $|V_{cb}|$.

\item [\it viii)]
The $O(\as^2)$ perturbative expansion for the $b\to c \ell\nu$ width 
has relative large coefficients   when $\overline{m}_c(3\GeV)$ is employed, while
the situation improves in case one uses $\overline{m}_c(2\GeV)$ or $m_c^{kin}(1\GeV)$, see the Appendix of \cite{NNLO}. We have briefly studied what happens with $\overline{m}_c(2\GeV)$:  the answer depends little on the way one computes it from $\overline{m}_c(3\GeV)=0.986(13)\GeV$.  Using three loop renormalization group evolution leads to $\overline{m}_c(2\GeV)=1.091(14)\GeV$, 
and in scenario {\bf D} one gets the results in the last row of Table \ref{tab:3}, with  $|V_{cb}|$ very close to the other fits.
Table \ref{tab:3} also reports results for a fit to  the
 charm mass expressed in the kinetic scheme. In this case 
 we employ scenario {\bf D} as in the default fit, and use a
  constraint on $m_c^{kin}(1\GeV)$ derived  from the $\overline m_c(3\GeV)$ determination of \cite{mcmb_karlsruhe}.  Using the translation formula given in   \cite{NNLO} we obtain
 $ m_c^{kin}(1\GeV)= 1.091\pm 0.020 \GeV$. The value of the BR and of $|V_{cb}|$  are consistent  with the results of the $\overline{\rm MS}$ scheme fits.
  
\end{itemize}

\begin{figure}
\begin{center}
\includegraphics[width=5.4cm]{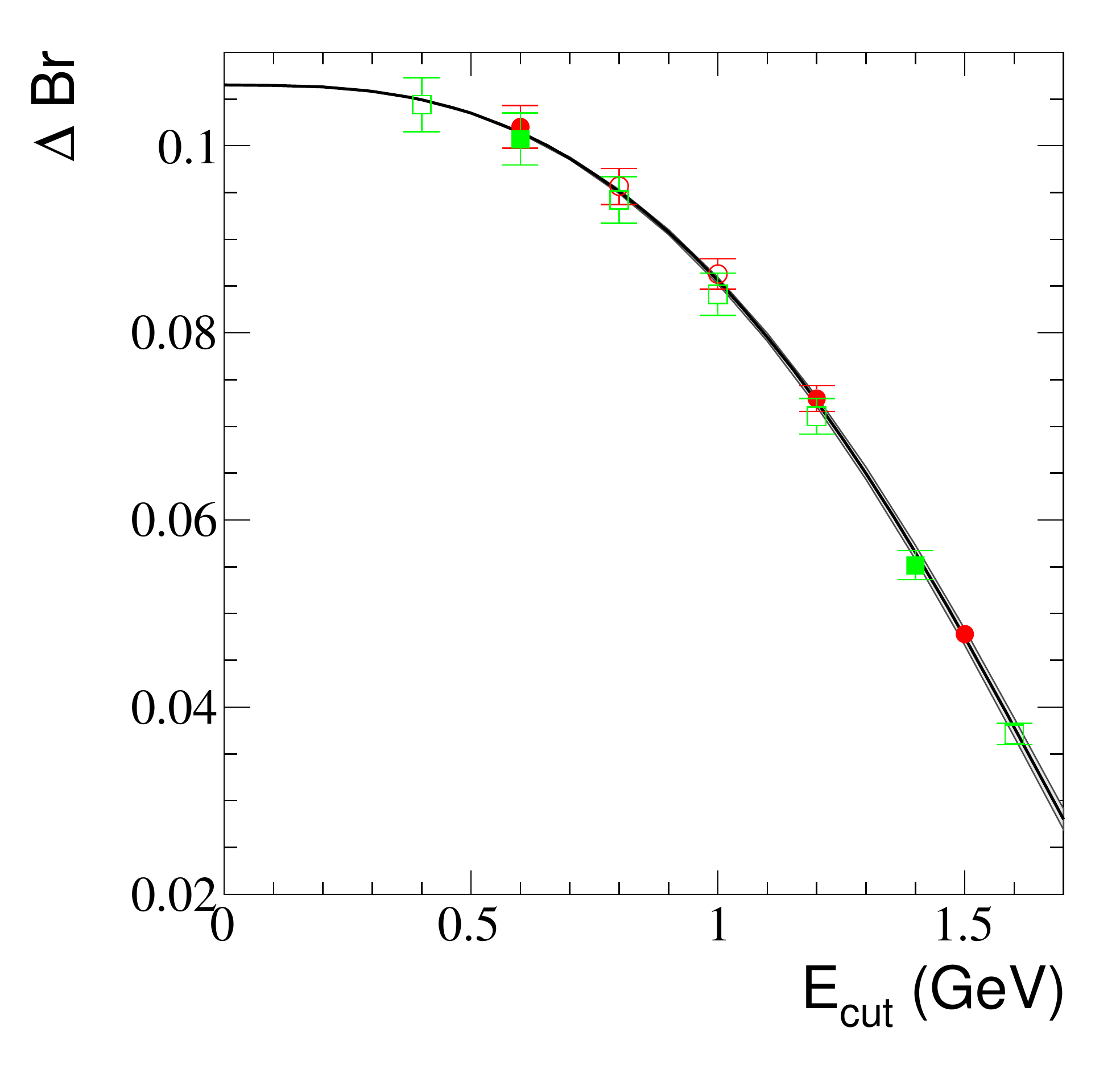}\ \ \
\includegraphics[width=5.4cm]{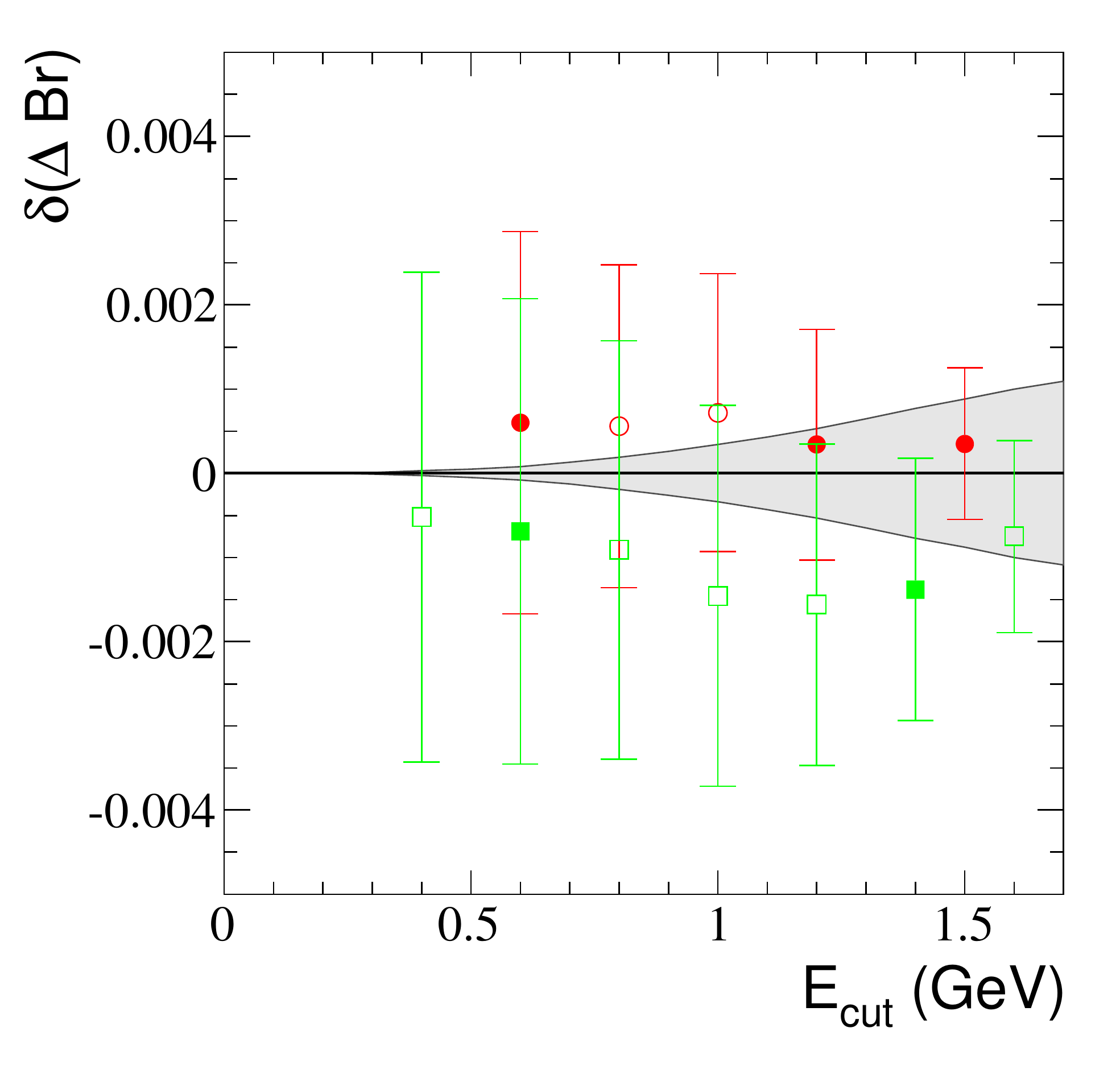}\\
\vspace{-2.5mm}
\includegraphics[width=5.4cm]{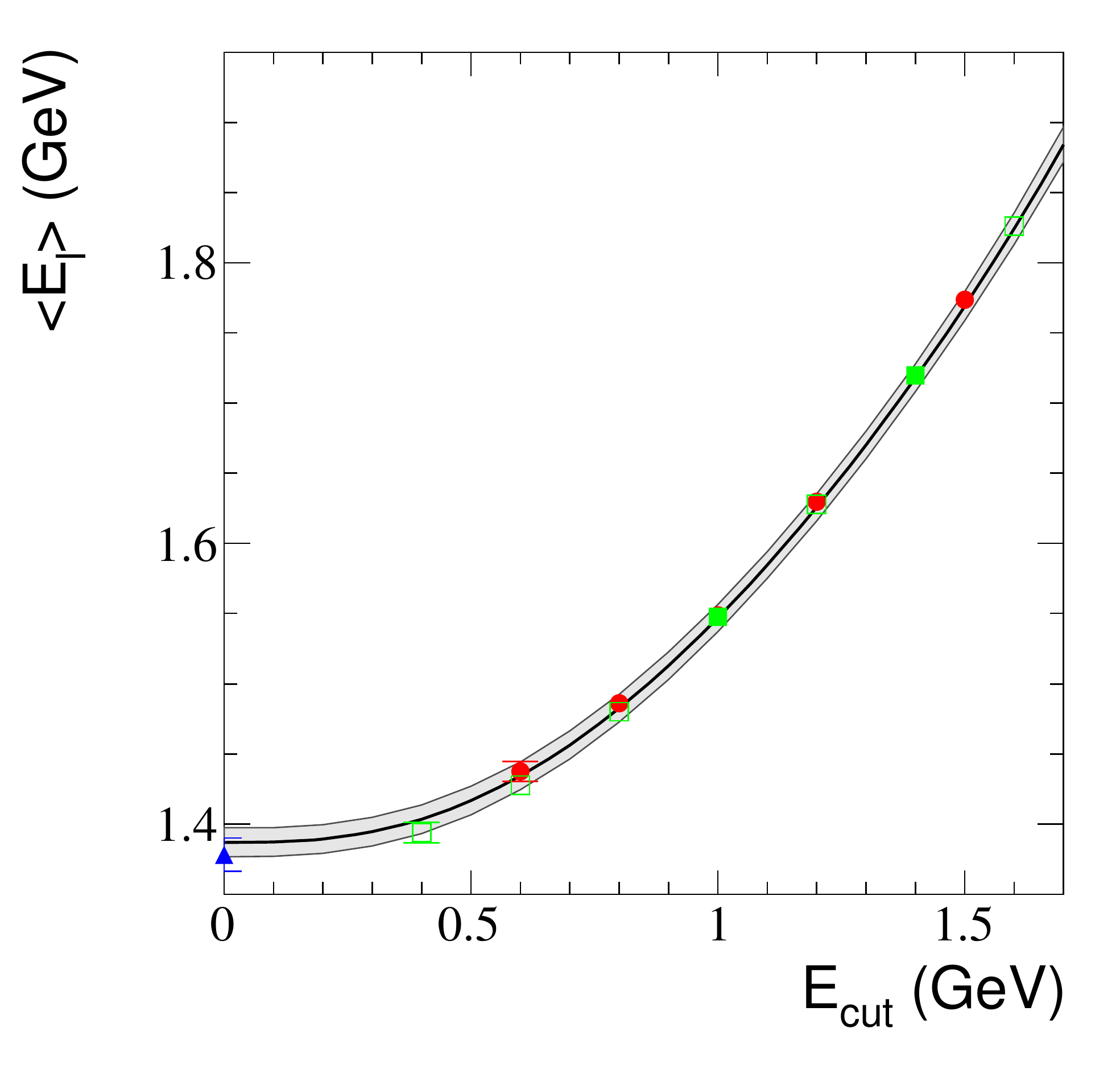}\ \ \
\includegraphics[width=5.4cm]{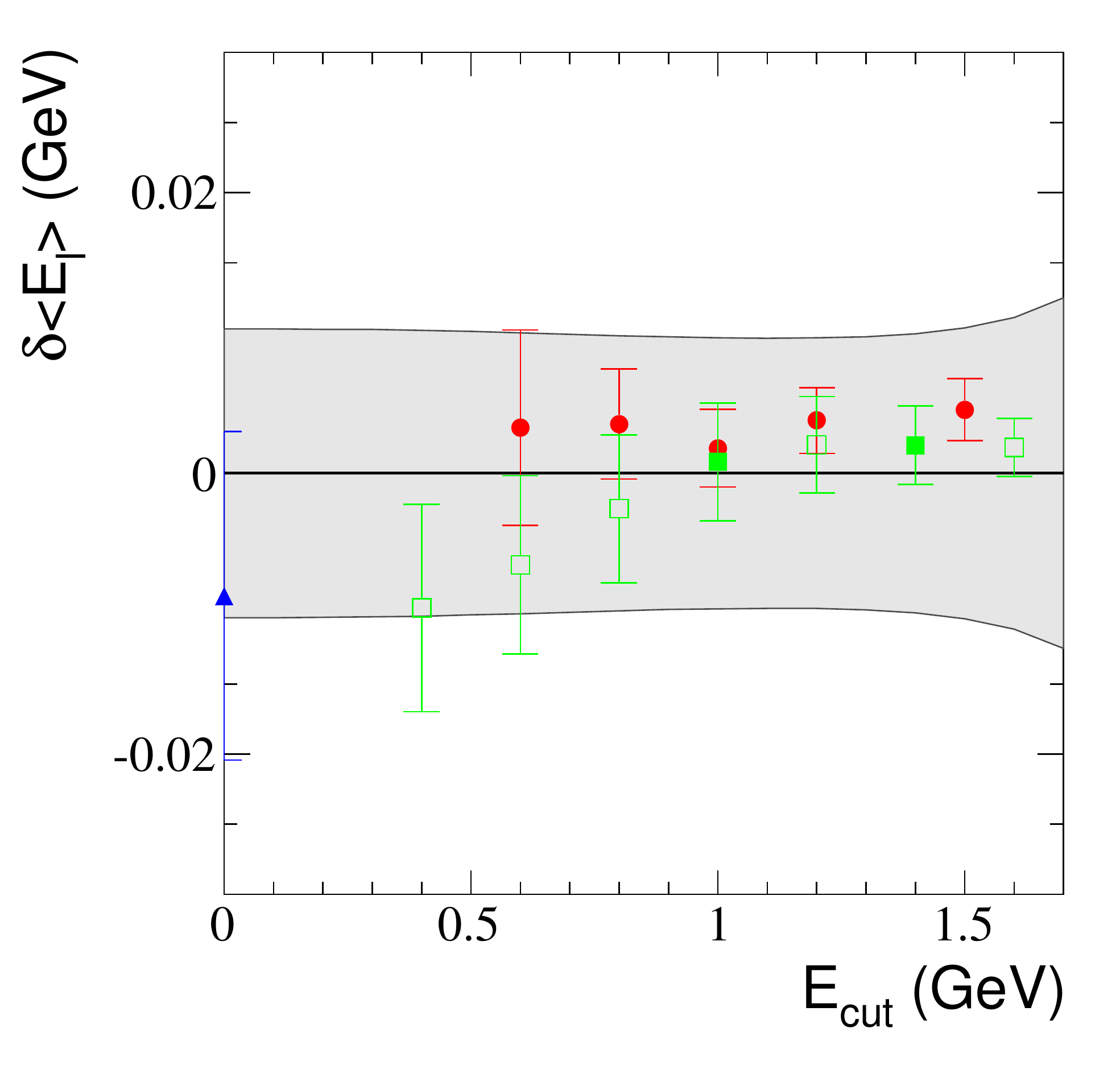}\\
\vspace{-2.5mm}
\includegraphics[width=5.4cm]{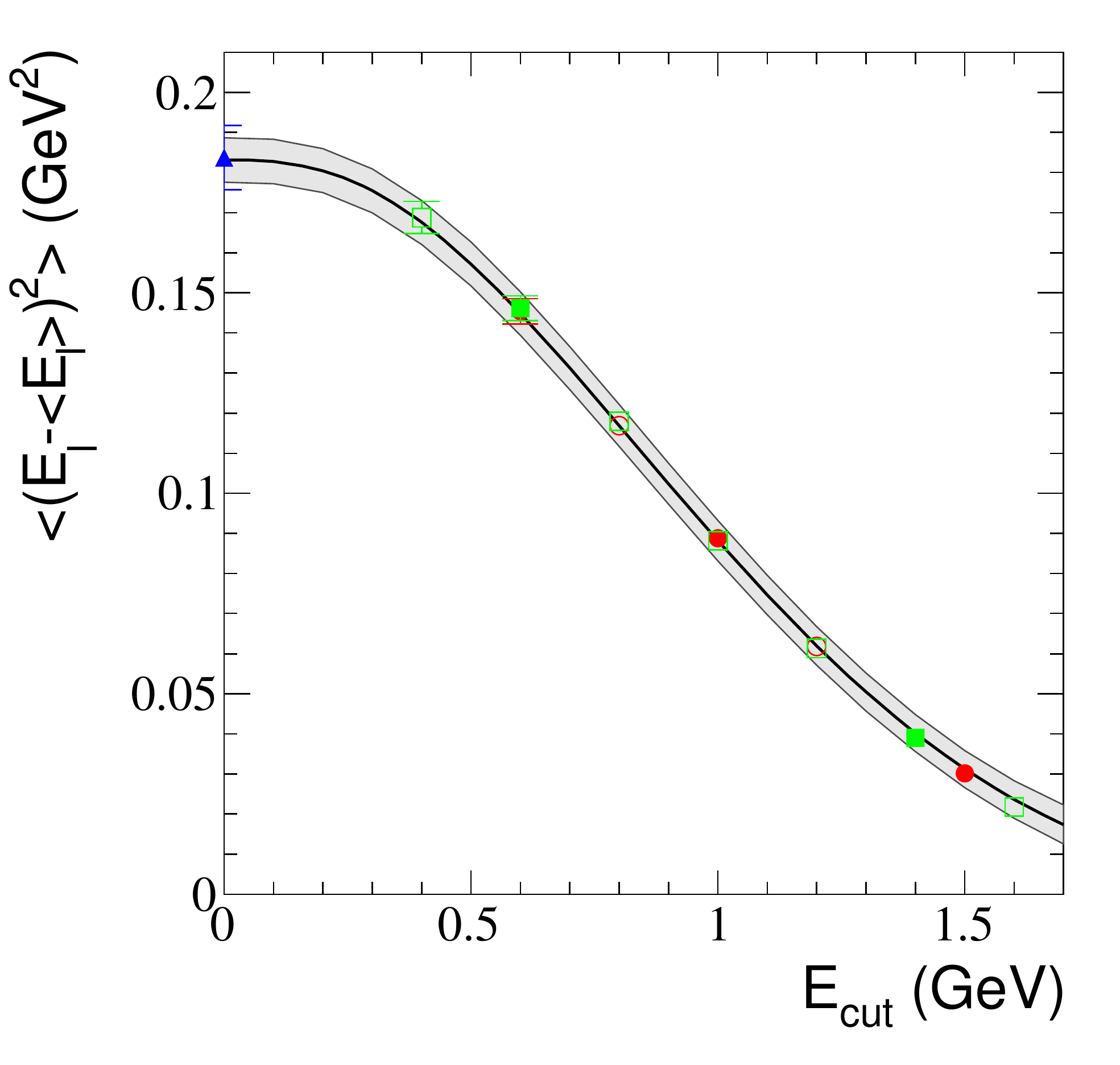} \ \ \
\includegraphics[width=5.4cm]{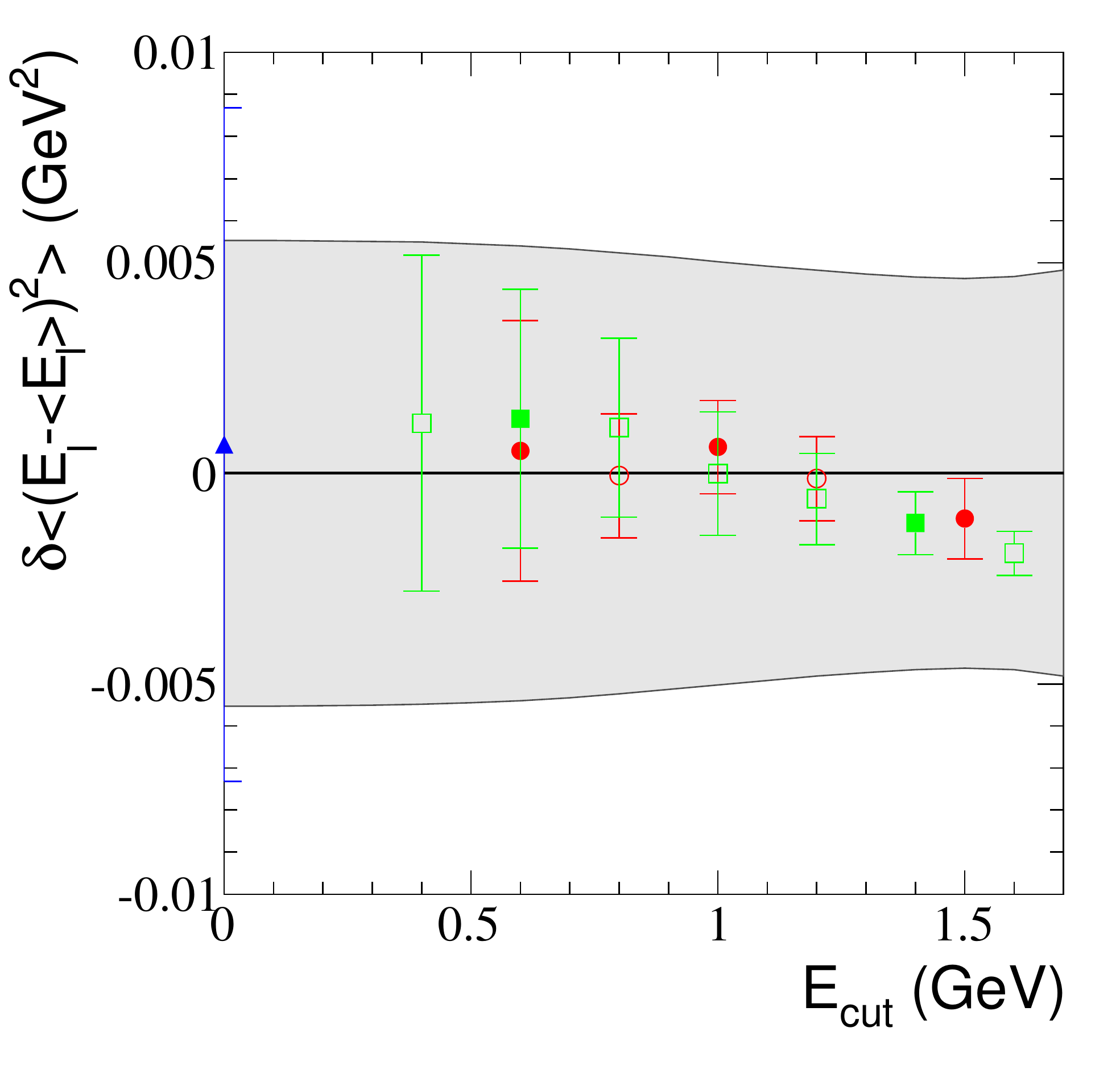}\\
\vspace{-2.5mm}
\includegraphics[width=5.4cm]{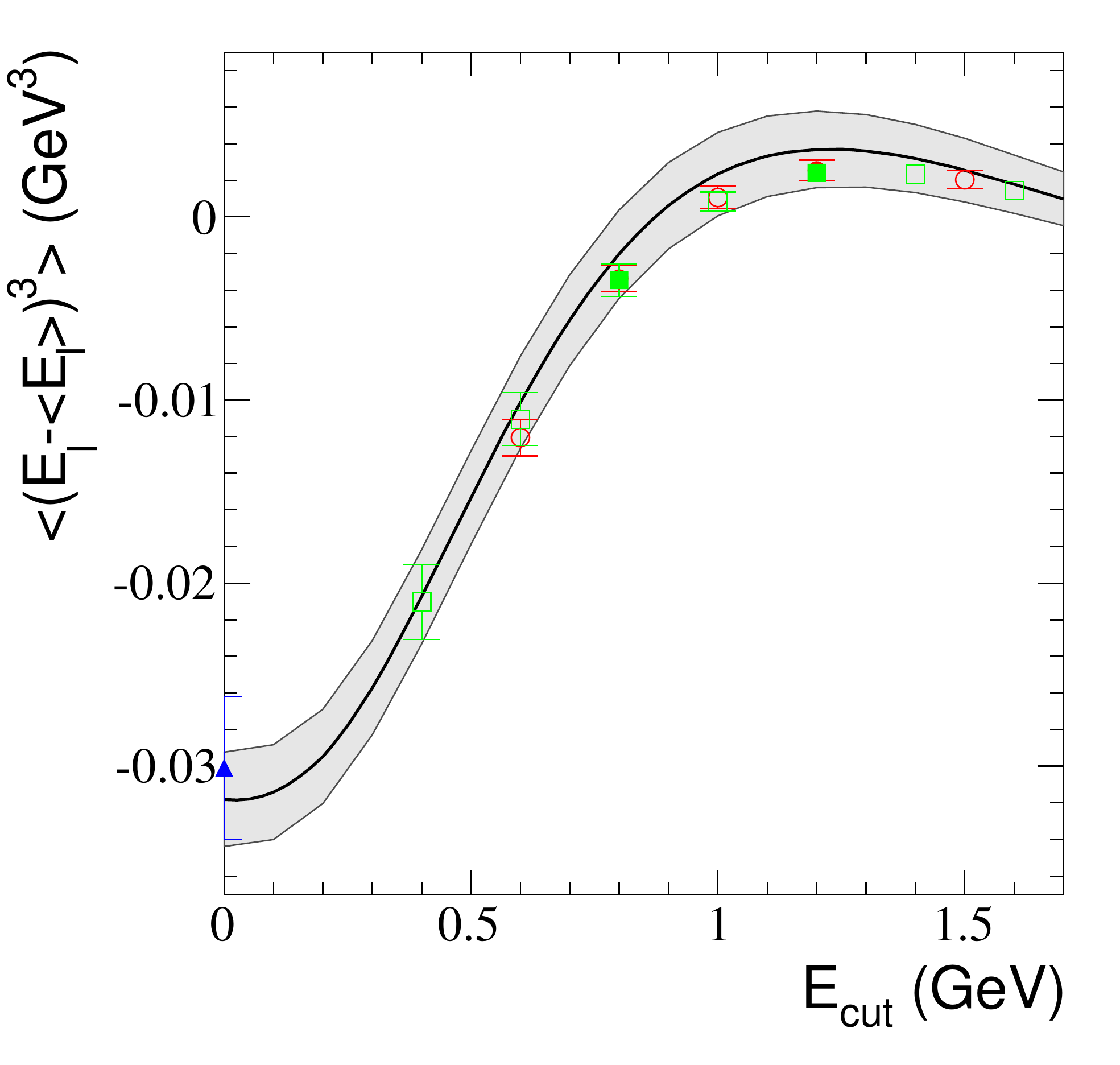}\ \ \
\includegraphics[width=5.4cm]{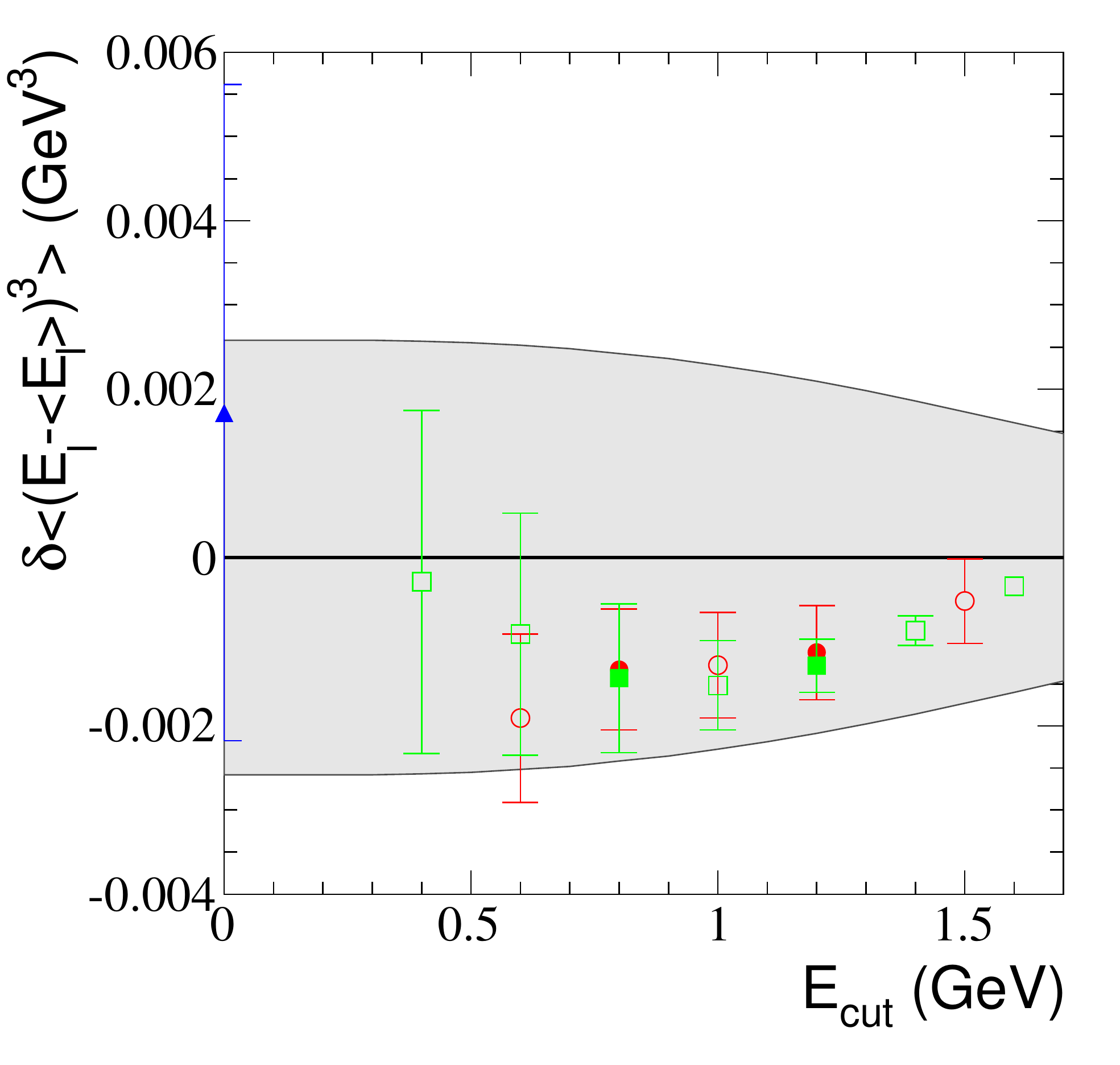}\vspace{-4mm}
\caption{\label{plots}  \sf Default fit predictions for $R^*, \ell_{1,2,3}$
compared with measured values in absolute terms (left) and  as deviations ($\delta$,right) from the predictions  as a function of $E_{cut}$.
 In all plots, the grey band is the theory prediction with total theory error. BaBar data are shown by circles, Belle by squares and other experiments (DELPHI, CDF, CLEO) by triangles. Filled symbols mean that the point was used in the fit. Open symbols are measurements that were not used in the fit.
}
\end{center}
\end{figure}

\begin{figure}
\begin{center}
\includegraphics[width=6.4cm]{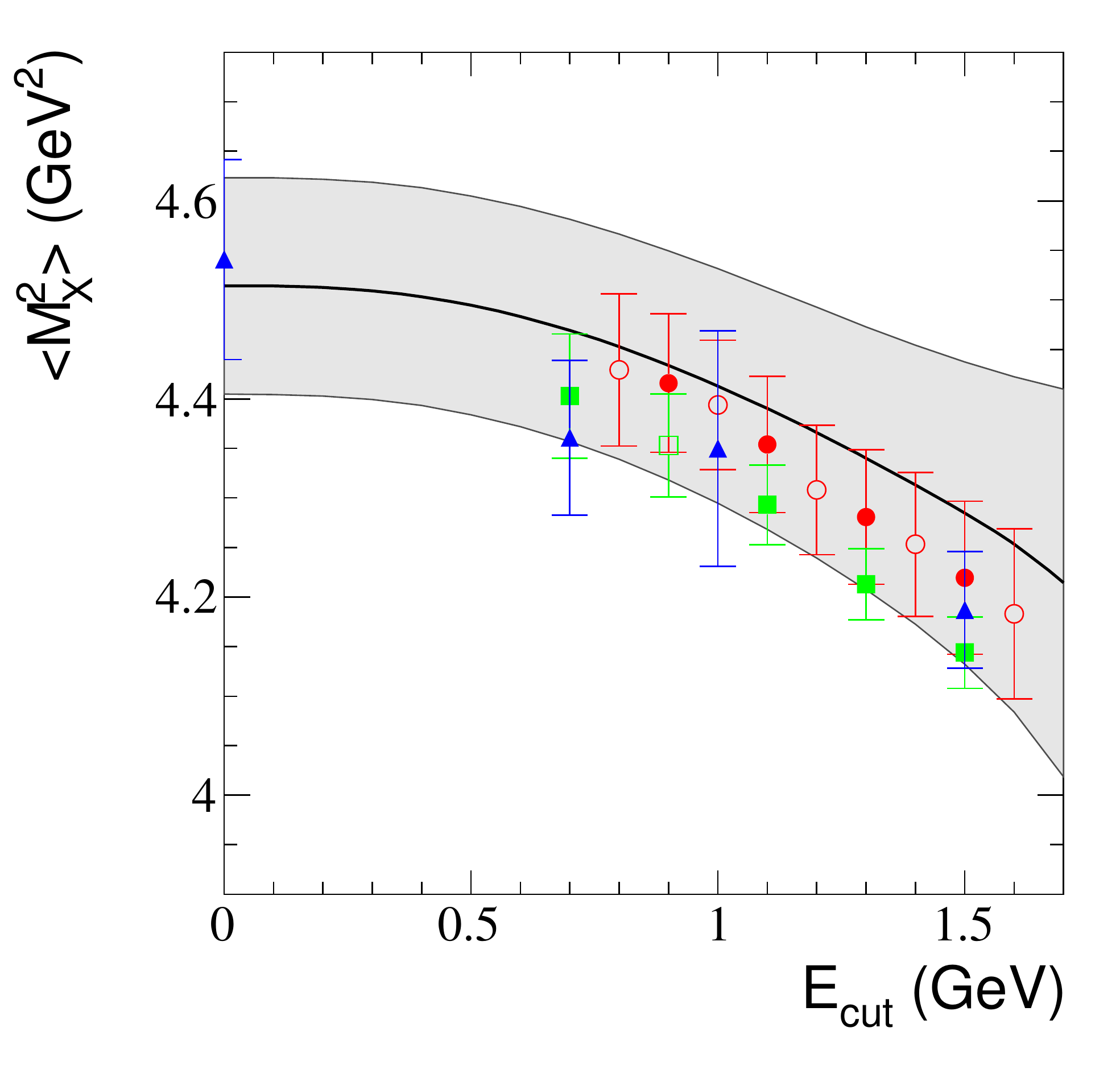}
\includegraphics[width=6.4cm]{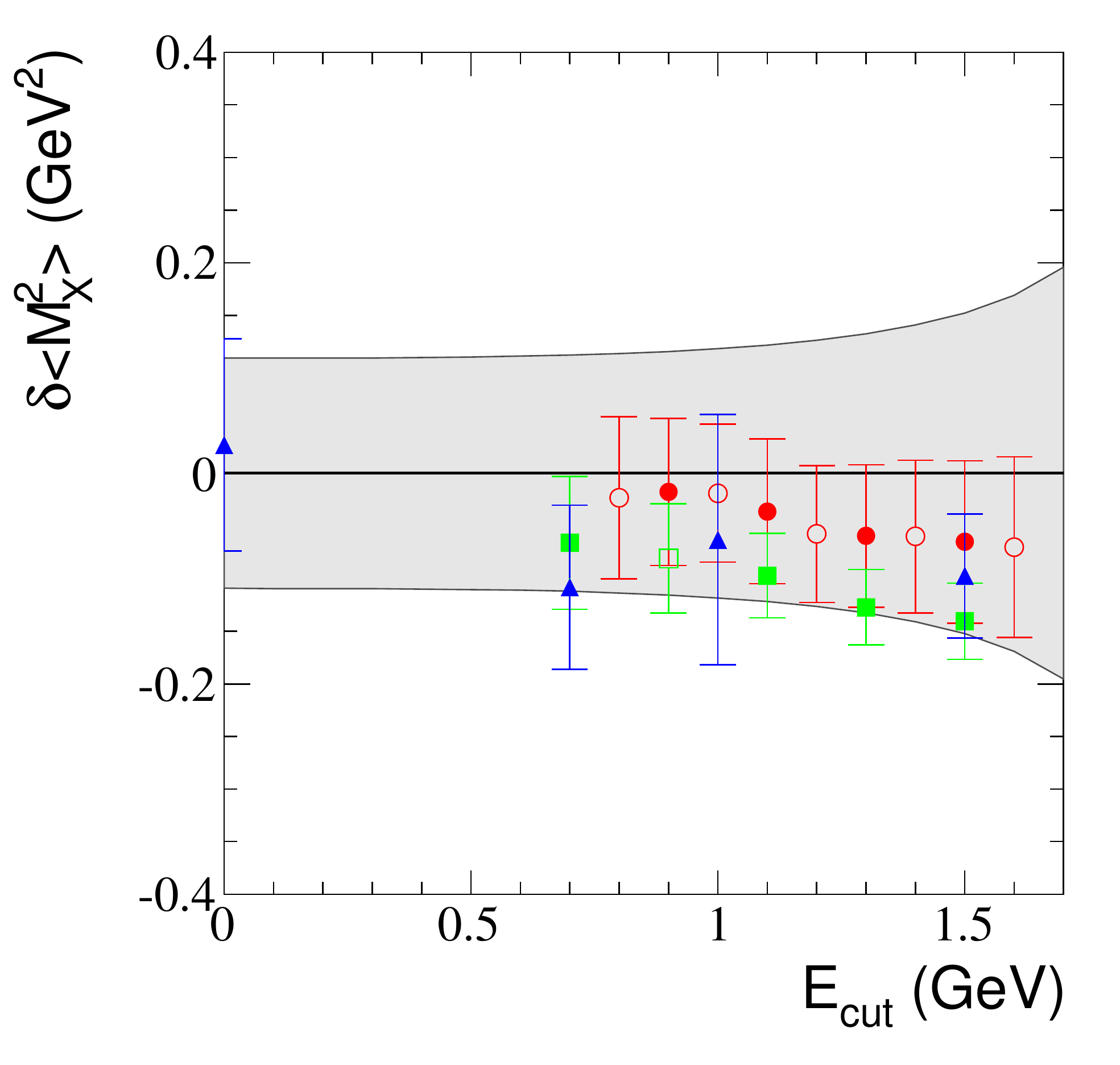} \\
\includegraphics[width=6.4cm]{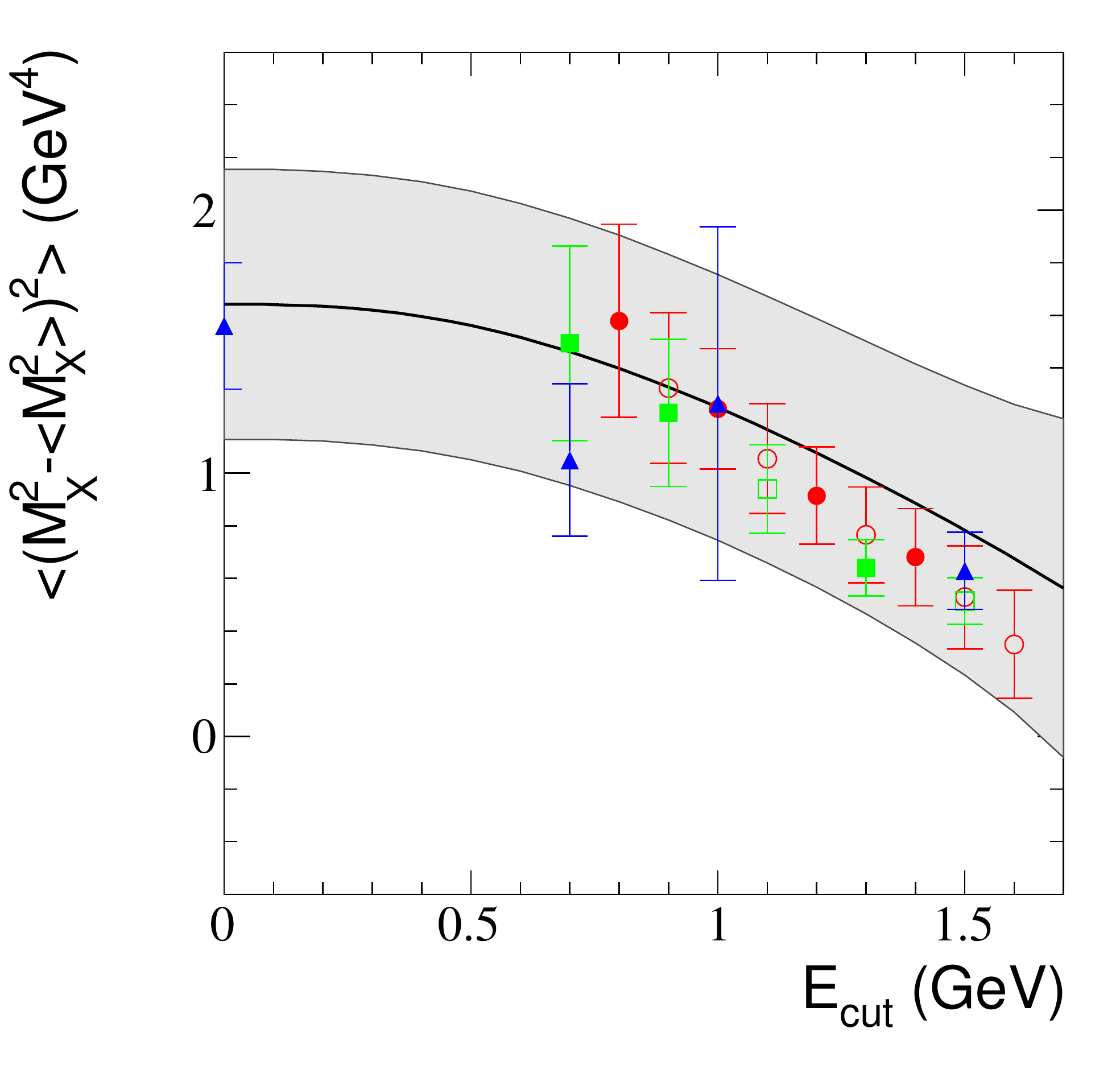}
\includegraphics[width=6.4cm]{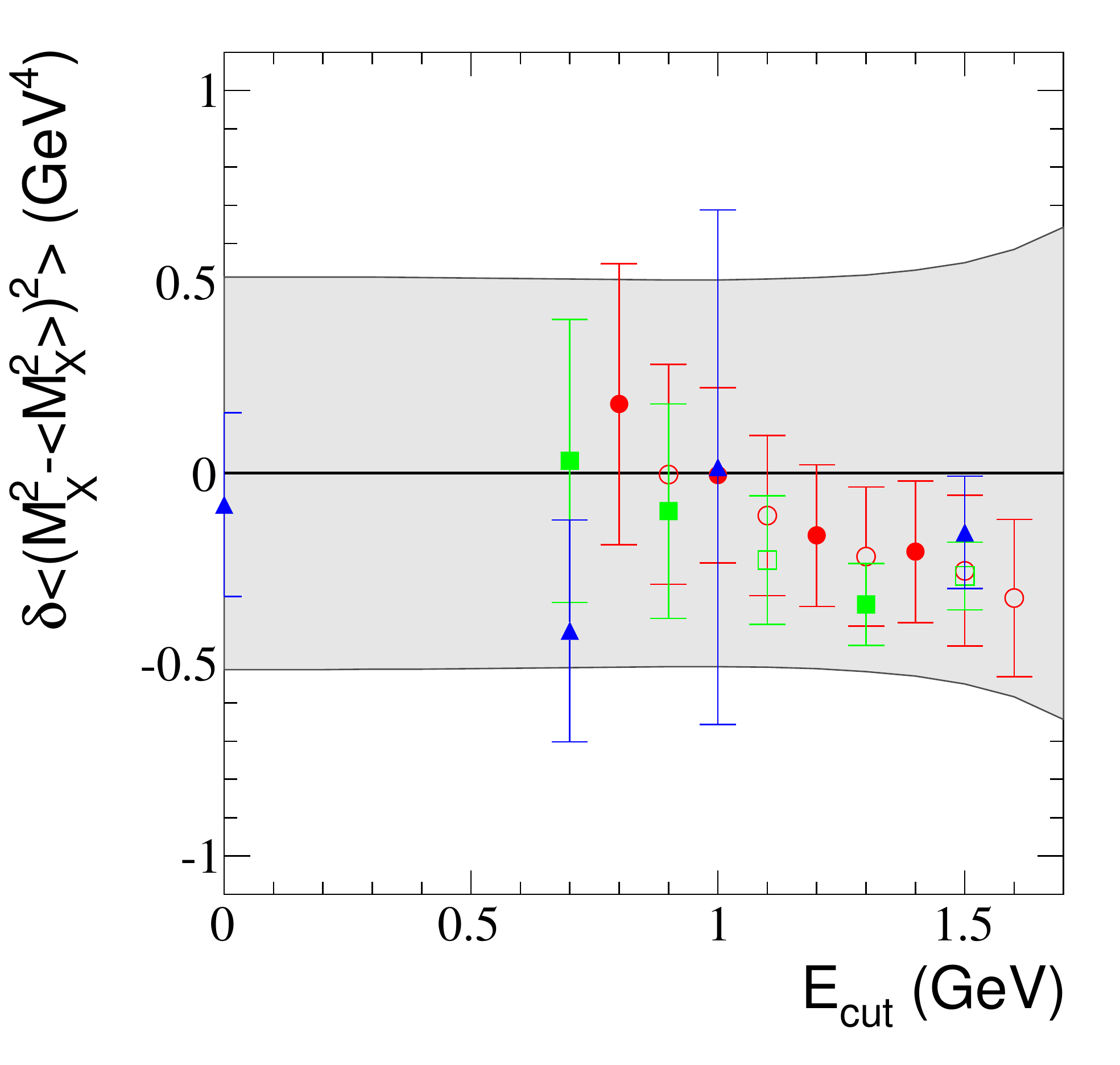}\\ 
\includegraphics[width=6.4cm]{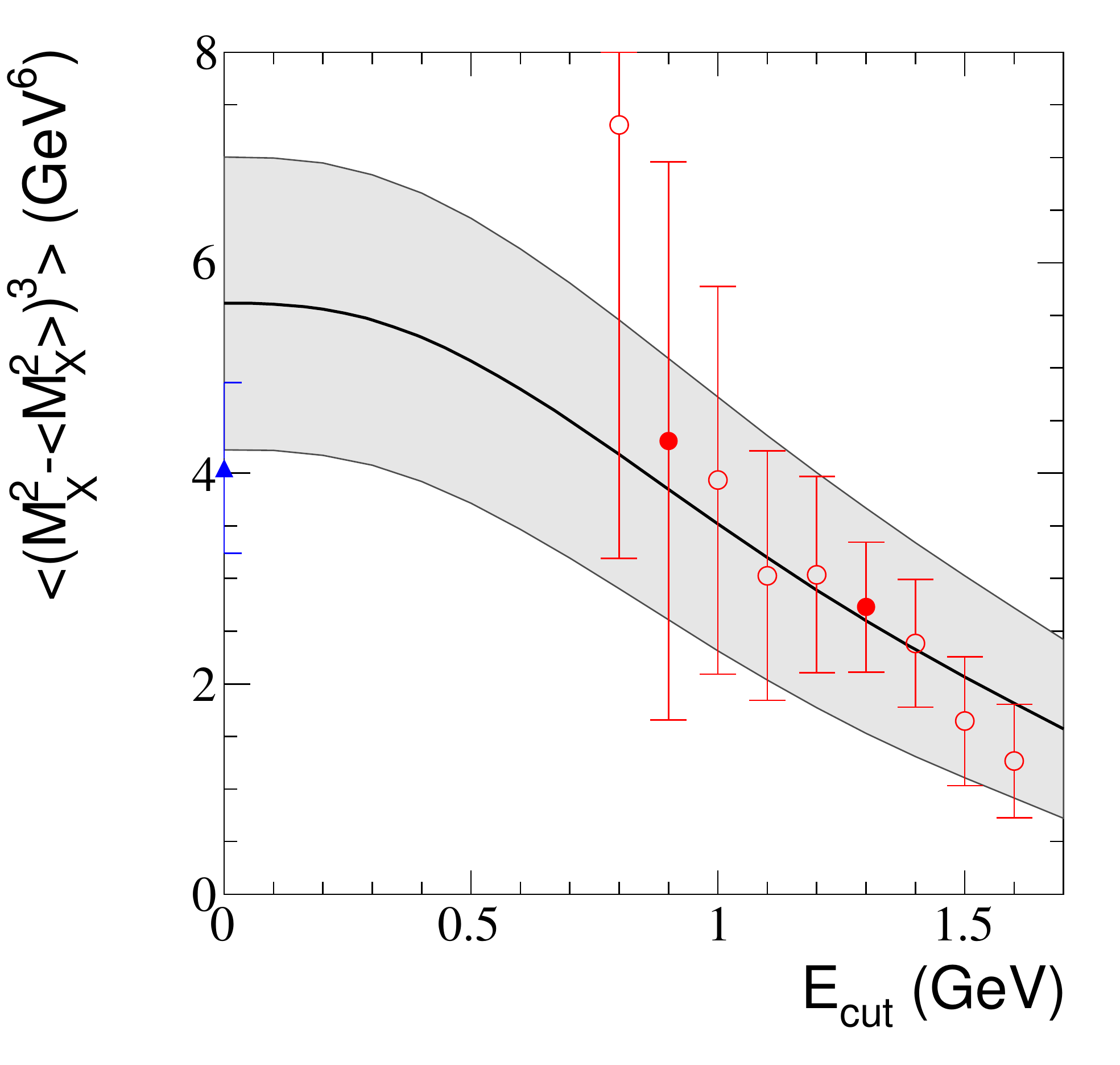}
\includegraphics[width=6.4cm]{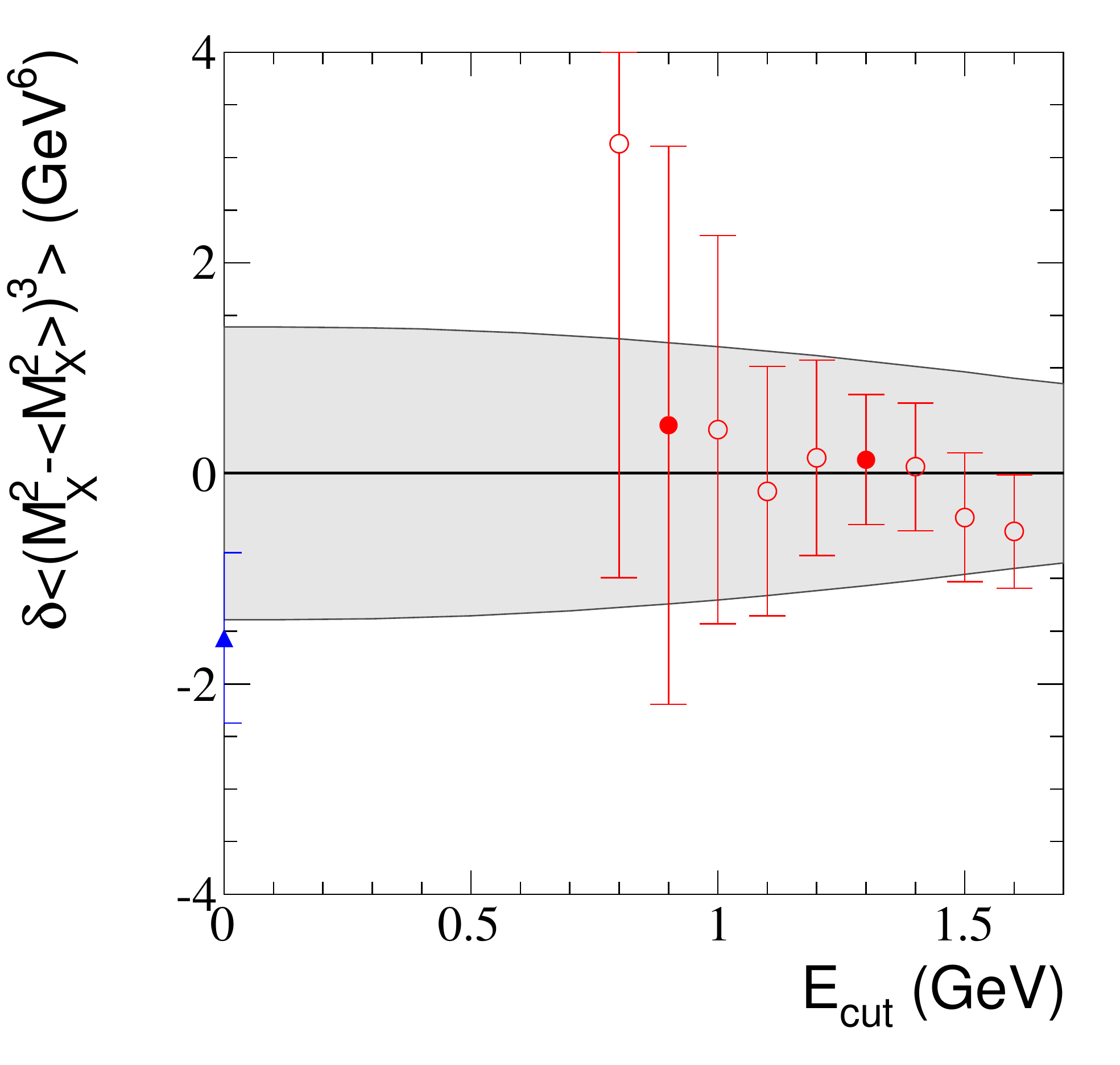}
\caption{\label{plots2}  \sf Same as in Fig.~\ref{plots} for $h_{1,2,3}$.}
\end{center}
\end{figure}

Because of strong correlations, the measurements listed in Table \ref{tab:1} are only a subset of all the measured moments.
In order to gain a visual appreciation of the quality of the default fit and to see how well it agrees also with  the measurements that are not included, we show  in Figs.~\ref{plots} and \ref{plots2} the leptonic and hadronic moments measurements compared with their theoretical prediction with theory uncertainty.  

\section{Implications of the default fit }

\subsection{Semileptonic phase space ratio }
It is particularly convenient to normalize the branching fraction of the rare  decays 
$B\to X_s \gamma$ and $B\to X_s \ell^+\ell^-$ 
to the semileptonic  one, BR$_{c\ell\nu}$.  In this context  the semileptonic phase space ratio
\be \label{phase2}
C = \left| \frac{V_{ub}}{V_{cb}} \right|^2 
\frac{\Gamma[\bar{B} \to X_c e \bar{\nu}]}{\Gamma[\bar{B} \to X_u e \bar{\nu}]}.
\ee
is usually factorized  \cite{gm,bobeth,ms,GG}. $C$ can be calculated using the OPE and the results of the fit to the semileptonic  moments. 
In principle, the $B\to X_u \ell \bar{\nu}$ width is also sensitive to Weak Annihilation
(WA) contributions, see e.g.\,\cite{GG}, which are poorly known but cancel out in the rare 
decay width. As far as the normalization of rare decays is concerned, WA effects 
can therefore be ignored. Even neglecting WA, however,   
the value of $C$  does depend on the scale at which the WA matrix element is assumed to vanish. In the following we follow \cite{GG} and use $\mu_{\scriptscriptstyle WA}=m_b/2$.

In Ref.~\cite{GG} $C$ was computed  in the kinetic scheme, based on the fits performed by HFAG  at that time. The result was 
$C=0.546\pm 0.016 (\rm pert) \pm 0.017 (\rm HQE)$, where the first uncertainty refers to 
higher order perturbative contributions, and the second is associated with the semileptonic fit. 
A different result, $C=0.582\pm 0.016$, was reported in \cite{1S}  in the 1S scheme, see also
the first paper of \cite{ms} for additional details. The central value has already been converted to our convention with $\mu_{\scriptscriptstyle WA}=m_b/2$.  
The discrepancy between these two values has now considerably reduced, 
as we will see in a moment. 

In order to compute $C$ with our default fit we need to adapt the calculation of \cite{GG} to
the use of a charm mass in the $\overline{\rm MS}$ scheme at $\mu=2$ or 3\,GeV.
Here we give the two corresponding approximate formulas, 
\bea
C&= &g\left(\rho\right)  \Big\{ 0.903 - 0.595 \,\delta_{\as} + 0.0405 \,\delta_b - 0.1137 \left(\overline m_c(2\GeV) - 1.05\GeV\right)\nonumber\\
&& \ \ \ \ \ \ - 0.0184\, \mug - 0.199\, \rd + 0.004\, \rls\Big\}, \label{Cmc2} \\
C&=& g\left(\rho\right)\Big\{ 0.849 - 0.92 \,\delta_{\as} + 0.0596 \,\delta_b - 0.2237 \left({\overline m}_c(3\GeV) - 1\GeV\right)\nonumber \\
&& \ \ \ \ \ \  - 0.0167 \,\mug - 0.203\, \rd + 0.004 \,\rls  \Big\},  \label{Cmc3}
\eea
where $g(\rho)= 1-8\rho+8\rho^3-\rho^4-12 \rho^2 \ln \rho$, $\rho=(m_c/m_b)^2$, 
$\delta_{\as}= \as(4.6\GeV)-0.22$, and $\delta_b= m_b-4.55\GeV$.
The approximate formulas reproduce the complete calculation to better than  0.4\% 
in the range  $4.45<m_b<4.65$GeV,  $0.9<\overline m_c(3\GeV)<1.1$GeV,
$0.95<\overline m_c(2\GeV)<1.15$GeV.  
Using our default fit we obtain 
\be
C=0.574\pm 0.008,\label{C1}
\ee
 with an additional  
$\sim 3\%$ theoretical error. The result is identical if we include   $m_b$ from \cite{mcmb_karlsruhe} in the fit. In scenario {\bf B} with both $m_{c,b}$ from \cite{mcmb_karlsruhe} we get $C=0.572(8)$. 
In comparison with \cite{GG}, the parametric uncertainty has reduced by  a factor 2, mostly 
thanks to the new charm mass constraint, and the central value has increased by about  1$\sigma$.
Our result is in good agreement  with that of  \cite{1S}.

One may worry that the perturbative series of $C$ in terms of $\overline{m}_c(3\GeV)$
has relatively large coefficients, as witnessed by the strong $\as$ dependence in (\ref{Cmc3}), and that
it likely involves large $O(\as^3)$ terms. In this respect, the situation is somewhat better if one 
employs the kinetic charm mass or $\overline{m}_c(2\GeV)$, see (\ref{Cmc2}). 
Evolving  $\overline{m}_c(3\GeV)=0.986(13)\GeV$ \cite{mcmb_karlsruhe}
down to  2\GeV, and using  $\overline{m}_c(2\GeV)=1.091(14)\GeV$ and $m_b^{kin}=4.533(32)\GeV$  as   constraints in a fit 
to  $\overline{m}_c(2\GeV)$ 
we get 
\be
C=0.566\pm 0.008,
\ee
again with additional 3\% theoretical uncertainty. This is  compatible with the result in Eq.~(\ref{C1}), and might be  preferred as a reference value.

Let us now address the reasons behind the shift of $C$ with respect to Ref.\,\cite{GG}, 
due only to the different set of inputs.
The fits we present in this paper, mostly because of the $m_c$ constraint, have significantly lower $m_b$ and $m_c$  than those used in \cite{GG}. Fig.~1 of Ref.\,\cite{GG} shows that  lowering the heavy quark masses one increases 
$C$ and that their new central values lead to $C\approx  0.57$. 
It should also be stressed that  in the absence of the $m_c$ constraint,  option A for the 
theoretical correlations leads to higher $m_b$ and $m_c$  and to artificially smaller errors than B,C,D \cite{Gambino:2011fz}. 
Therefore previous fits in the kinetic scheme preferred a lower $C$ and underestimated its uncertainty. The analogous effect on $|V_{cb}|$ was  negligible.
This is also the main reason for the discrepancy between the values of C in \cite{GG} and \cite{1S}.  There are other differences between these two analyses (experimental inputs, scheme, etc.), but this is the most important component.

Of course, the factor $C$ is just one component of the calculation of rare $B$ decays. 
What actually enters inclusive $B$ decays is ${\rm BR}_{c\ell\nu}/C$, and there are additional power-suppressed corrections that affect  the   
width and depend on the same parameters that determine $C$, as well as 
charm loops in the perturbative corrections. As precision  increases,  
it is no longer obvious that the factorization of $C$ is still advantageous, except as a bookkeeping device. Nevertheless, 
the results of our fit, with the complete correlation matrix, are all one needs
for a careful analysis of the parametric uncertainty in those cases. 

Finally, it is worth reminding that, in the case of $B\to X_s\gamma$, it is often necessary to  
extrapolate measurements performed with a cut on the photon 
energy higher than about 1.6\GeV\ to lower photon energies, where the local OPE is expected to work better. This extrapolation  can be performed using different techniques \cite{extra}, but it crucially depends on precise HQE parameters, namely on the results of semileptonic fits.

\subsection{Local contributions to the zero-recoil sum rule}
The $B\to D^* \ell \nu$ form factor at zero recoil can be estimated using 
heavy quark sum rules \cite{sumrules,long}. The 
 form-factor $ {\cal F}(1)$
is obtained by separating 
the elastic $B\to D^*$ transition contribution from the total inelastic 
transition at zero-recoil: 
\be
I_0(\varepsilon_{M})= {\cal F}^2(1)+I_{\rm inel}(\varepsilon_{M}), 
\label{84}
\ee
where $I_{\rm inel}(\varepsilon_{M})$ is related to the sum of the
differential decay probabilities into the excited states with mass up to
$M_{D^*}\!+\!\varepsilon_{M}$ in the zero recoil kinematics. 

The OPE allows us to calculate the amplitude 
$I_0(\varepsilon_{M})$  in the short-distance expansion provided
$|\varepsilon|$ is sufficiently large compared to   the ordinary 
hadronic mass scale. Setting $\varepsilon=\mu^{kin}$ 
\be
I_0(\mu^{kin})= \xi_A^{pert}(\mu^{kin}) - \Delta_{1/m^2} -\Delta_{1/m^3}+...,
\ee
where the ellipses stand for  higher order contributions and $ \xi_A^{pert}$ represents a perturbative contribution. The latter was computed in Ref.~\cite{long} to $O(\as^2)$: 
$ \sqrt{\xi_A^{pert}}(0.75\GeV)=0.98\pm 0.01$.
The leading power contributions to $I_0$ were calculated in
Refs.~\cite{sumrules} to order $1/m_Q^2$ and  in
Ref.~\cite{rev} to order $1/m_Q^3$ and read
\bea
\Delta_{1/m^2} &=& \frac{\mu_G^2}{3m_c^2} +
\frac{\mu_\pi^2 \!-\!\mu_G^2 }{4}
\left(\frac{1}{m_c^2}+\frac{2}{3m_cm_b}+\frac{1}{m_b^2}
\right),  \nonumber\\
\Delta_{1/m^3} &=&
\frac{\rho_D^3 - \frac{1}{3}
\rho_{LS}^3}{4m_c^3}\;+\;
\frac{1}{12m_b}\left(\frac{1}{m_c^2}+\frac{1}{m_c m_b} +\frac{3}{m_b^2}\right)
\,(\rho_D^3 +
\rho_{LS}^3)\,.
\nonumber
\eea
In the kinetic scheme the nonperturbative parameters  $\mu_\pi^2$, $\mu_G^2$, $ \rho_D^3$ and
$\rho_{LS}^3$ all depend on the hard Wilsonian cutoff $\mu^{kin}$.  Since the heavy quark expansion of $I_0$ involves inverse powers of $m_c$, the cutoff must satisfy 
$\mu^{kin}\ll 2 m_c$ and it is better to have it 
lower than 1\GeV.  
We therefore take the values of the OPE parameters
extracted from our default  fit with $m_c$ in the kinetic scheme, evolve them down to $\mu^{kin}=0.75\GeV$ and find
\be
\Delta_{1/m^2}= 0.084\pm0.017,  \qquad \Delta_{1/m^3}= 0.021\pm 0.008\, ,
\ee
and for the sum
\be
  \Delta_{1/m^2}+\Delta_{1/m^3} = 0.104\pm 0.023 \, .
\label{110a}
\ee
If we use scenario {\bf B} we get a slightly larger value $0.111\pm 0.016$,  closer to the preliminary result $0.118\pm 0.015 $ given in \cite{long} and obtained using that scenario.
The quantity in (\ref{110a}) is an important ingredient of the heavy quark sum rule estimate 
of  $ {\cal F}(1)$: the present update does not affect significantly the results of \cite{long}.

\subsection{Impact on inclusive $|V_{ub}|$ determination}
In order to get a rough estimate of the precision 
that can be reached applying the results of our default fit to the semileptonic $B\to X_u \ell\nu$ analyses, we use the information given in the kinetic scheme analysis of Ref.~\cite{GGOU}.
Recent experimental studies have considered  lower cutoffs 
on the lepton momentum as low as 1\GeV, which is very close to the total decay width
and for which a local OPE description is perfectly adequate.
Therefore, we take Eq.~(45) of Ref.~\cite{GGOU} and compute  the
parametric uncertainty on the total $B\to X_u \ell \nu$ width using the results of the default fit. 
We obtain a 2.2\% error that translates into a 1.1\% parametric uncertainty on $|V_{ub}|$.
A slightly smaller uncertainty, 1.8\%, is obtained if one employs a fit that also includes the 
constraint on $m_b$ from \cite{mcmb_karlsruhe}. Since the uncertainty is dominated by 
that of $m_b$, better determinations of this mass will result in further reduction of the parametric error.

In summary, the parametric uncertainty on the total $B\to X_u \ell \nu$ width is about 2\%
and can be reduced in the future; it will have to be considered together with 
a $\gsim 1\%$ theoretical uncertainty. As mentioned above this should essentially hold 
even for a lower cut on the lepton momentum around 1\GeV\ is applied. On the other hand, 
for higher cuts the local OPE is no longer sufficient and the sensitivity to $m_b$ gets 
stronger: these analyses will benefit of a better determination of $m_b$ even more.

\section{Summary}

In this paper we have reassessed the whole strategy of global  fits to the semileptonic 
moments. We have shown that the results depend sensitively not only on the estimate of the 
theoretical uncertainties, but also on 
the assumptions about  their correlation. We have studied the impact of precise determinations
of the heavy quark masses from independent data on the global fits and shown that
their use leads to more precise results, which depend less on the assumptions on the 
theoretical correlations.  Using a determination of $m_c$ with $ 13$MeV uncertainty we 
were able to determine $m_b$ within about 20 MeV, in good agreement and competitive 
with some of the most precise $m_b$ determinations.
 In the absence of external constraints on the heavy quark masses, 
the semileptonic moments  determine their difference with a 20MeV uncertainty, see Eq.~(\ref{massdiff}). This is a robust NNLO relation that we find stable against theoretical assumptions.

The value of  $|V_{cb}|$ that  we obtain is  higher than in previous analyses, but compatible with 
the prediction of a global CKM  fit in the Standard Model \cite{UTfit}; it has  a total 2\% accuracy, which is dominated by theoretical errors.
  We have also studied the impact of the new fits on the calculation of the
semileptonic phase space ratio $C$, on the power corrections to the zero-recoil sum rule, and on the extraction of $|V_{ub}|$.

Theoretical uncertainties are the major obstacle to an accurate determination
of $|V_{cb}|$ using inclusive semileptonic $B$ decays.
Our theoretical errors are essentially  determined by a conservative estimate of the dominant sources of higher order corrections. There are therefore good prospects for
improvement, related to the completion of the calculation of $O(\as \mug/m_b^2)$ and to 
the inclusion of the $O(1/m_Q^{4,5})$ corrections. 
A more significant reduction of the error on $|V_{cb}|$ will also require a calculation of perturbative corrections to the coefficient of the Darwin operator. 
In contrast, the experimental situation is much better at the moment and will further 
improve at Belle-II.

\section*{Acknowledgements}
PG is grateful to Lorenzo Magnea and to the late Kolya Uraltsev
for useful discussions, and to Mikolaj Misiak and  Matthias Steinhauser for carefully reading 
a preliminary version of this paper. This work is supported in part by  MIUR under contract 2010YJ2NYW$\_$006.

\end{document}